\documentstyle[12pt,epsfig]{article}

\newcommand{\be}{\begin{eqnarray}}
\newcommand{\ee}{\end{eqnarray}}

\def\nue{{\nu_e}}

\newcommand{\dm}{\mbox{$\Delta m_{21}^2$~}}

\newcommand{\kl}{\mbox{KamLAND~}}

\begin{document}
%%%%%%%%%%%%%%%%%%%

\begin{flushright}
\end{flushright}

\begin{center}
{\Large \bf Solar neutrinos and 1-3 leptonic mixing
}
\vspace{.5in}

Srubabati Goswami$^{1}$,
%\footnote{e-mail: sruba@mri.ernet.in},
Alexei Yu. Smirnov$^{2,3}$,
\vskip .5cm

$^1${\small {\it Harish-Chandra Research Institute, Chhatnag Road, Jhusi,
Allahabad  211 019, India}},\\
$^2${\small{\it The Abdus Salam International Centre for Theoretical 
Physics,
I-34100, Trieste, Italy}}\\
$^3${\small{\it Institute for Nuclear Research, Russian Academy of
Scienses, Moscow, Russia}}\\
\vskip 1in

\end{center}

\begin{abstract}
Effects  of the 1-3 leptonic mixing on  the solar neutrino observables are 
studied  and the signatures of non-zero $\theta_{13}$ are identified. 
For this we have re-derived the  formula for $3\nu$-survival probability 
including all relevant corrections and constructed 
the iso-contours of observables in the 
$\sin^2 \theta_{12} - \sin^2 \theta_{13}$ plane. 
Analysis of the solar neutrino data  gives   
$\sin^2\theta_{13} = 0.007^{+ 0.080}_{-0.007}$ ($90\%$ C.L.) for 
$\Delta m^2 = 8 \cdot 10^{-5}$ eV$^2$.  
The combination of the ratio CC/NC at SNO and  gallium production rate 
selects $\sin^2\theta_{13} = 0.017 \pm 0.026$  ($1\sigma$). 
The global fit of all oscillation data leads 
to zero best value of $\sin^2 \theta_{13}$. 
The sensitivity ($1\sigma$ error) of future solar  neutrino studies to 
$\sin^2 \theta_{13}$  can be improved  
down to 0.01 - 0.02 by  precise 
measurements of the  pp-neutrino flux and the CC/NC ratio as well as 
spectrum distortion at high ($E > 4$ MeV) energies.  
Combination of experimental results sensitive to the 
low and high energy  parts of the solar neutrino spectrum   
resolves the degeneracy of angles $\theta_{13}$ and  $\theta_{12}$. 
Comparison of $\sin^2 \theta_{13}$ as well as $\sin^2 \theta_{12}$ measured 
in the solar neutrinos and in the 
reactor/accelerator experiments may reveal new effects 
which can not be seen otherwise. 

\end{abstract}
%%\pacs{14.60.Pq}   

%%\maketitle   

%%%%%%%%%%%%%%%%%%%%%%
\section{Introduction}
%%%%%%%%%%%%%%%%%%%%%%

Determination of the 1-3 mixing parameterized by the angle $\theta_{13}$  
is of great importance for phenomenology,  
for future experimental programs, and eventually,  for 
understanding the underlying physics.

At present,  for  $\sin^2 \theta_{13}$ we only have the upper bounds.  
The CHOOZ~\cite{chooz} reactor experiment (see also PaloVerde \cite{palo})  
gives 
\be
\sin^2 \theta_{13} \leq 0.036, ~~~  90\%~~ {\rm C.L.}  
\ee 
for $\Delta m^2_{13} = 2.5 \cdot 10^{-3}$ eV$^2$.    
The recent global fit of the oscillation results which includes also 
the solar salt phase II SNO, \kl and atmospheric neutrino data leads 
to~\cite{lisi05}
%~\cite{john,kl766us,valen04} 
\be
{\sin^2 \theta_{13} \leq 0.026 ~(0.047), ~~~ 90\% ~ (3\sigma) ~~{\rm C.L.}.
}
\label{limm}
\ee
According to \cite{strumia} the 90\% C.L. bound is $\sin^2 \theta_{13} 
\leq 0.024$.   
The limit (\ref{limm}) is slightly better than the earlier 
(before SNO salt phase II) limits ~\cite{john,kl766us,valen04} $\sin^2 
\theta_{13} \leq 0.031 ~(0.055)$.   
In \cite{concha04} even stronger  $3\sigma$ upper bound is quoted:  
$\sin^2 \theta_{13} \leq  0.041$. 

The combined fit of the CHOOZ, K2K and atmospheric neutrino results 
(without solar ones) gives~\cite{lisi05}    
\be
\sin^2 \theta_{13} \leq 0.032 ~(0.060), 
~~~ 90\%~  (3\sigma) ~~{\rm C.L.}.
\label{limglob}
\ee
(See also \cite{sgnu04}.) 
Comparison of the bounds (\ref{limm})  and (\ref{limglob}) 
shows that effect of inclusion of the solar neutrino data in the global 
fit is weak but not negligible:  the bounds improve by $\sim 25\%$.\\ 
%certain relevance of the present solar neutrino  data
%in restriction of 1-3 mixing. 

Future long baseline  experiments
are geared towards precision measurement of $\theta_{13}$. 
MINOS~\cite{minos} and  CERN to Gran-Sasso~\cite{cngs} alone 
will be able to only slightly improve the limit (\ref{limm})  
down to 
\be 
\sin^2 \theta_{13} \leq 0.015, ~~~90 \%~~ {\rm C.L.}.  
\label{th13lbl}
\ee
J-PARC accelerator experiment~\cite{jparc} will 
have substantially  higher sensitivity: 
\be
\sin^2\theta_{13} \leq 0.0015, ~~~~   90 \%~~ {\rm C.L.},  
\label{jparc}
\ee
%at $\Delta m^2 = 3\times 10^{-3}$ eV$^2$. 
but when all correlations and degeneracies of parameters 
are taken into account the estimated sensitivity reduces 
to $\sin^2\theta_{13} \approx 0.004~ (90 \%~~ {\rm C.L.})$~\cite{Huber}. 

The  Double-CHOOZ reactor experiment~\cite{2chooz}   will be able to 
improve the  bound down to  
\be
\sin^2 \theta_{13} = 0.008,  ~~~90 \%~~ {\rm C.L.} 
\label{dchooz}
\ee
for $\Delta m^2 =  2.4 \times 10^{-3}$ eV$^2$.  
It provides a clean channel of measurement of $\theta_{13}$ complementary 
to accelerator measurements~\cite{yasuda},\cite{Huber}. 

Detection of neutrinos from the Galactic supernova  can, in 
principle, probe $\sin^2\theta_{13}$ down to $10^{-5} - 10^{-6}$. 
The problem here is large uncertainties in the original neutrino fluxes.  
Essentially one can find  whether $\sin^2\theta_{13}$ is larger or smaller 
than $10^{-3}$. Both the upper and the lower bounds on $\sin^2\theta_{13}$ 
may be established if $\sin^2\theta_{13}$ is in the range 
$10^{-5} - 10^{-3}$ \cite{Lunardini:2003eh,Fogli:2005ad}. 

Ultimate sensitivity, $\sin^2\theta_{13} = 10^{-4} - 10^{-5}$, 
can be achieved at the neutrino factories~\cite{nufac}.\\

Influence of the 1-3 mixing on the detected solar neutrino fluxes has been 
considered in a number of publications before 
\cite{bari1} - \cite{lim03}, \cite{valen04}.  
The present solar neutrino data are not precise enough  
to really probe $\sin^2\theta_{13}$ in the allowed range 
(\ref{limm},\ref{limglob}), but on the other hand the bounds obtained 
using the solar data only are not much weaker than those from 
the reactor and atmospheric neutrino results.   
Indeed,  recent analysis of the solar neutrino data alone \cite{lisi05} 
gives 
\be
\sin^2 \theta_{13} \leq 0.038 ~(0.062), ~~~ 90\% ~ (2\sigma) ~~{\rm C.L.},  
\label{limsol}
\ee
and very weak $3\sigma$ bound. 
%%which is much weaker than the bound (\ref{limglob}). 
%In~\cite{valen04} it is found that 
Inclusion of the latest KamLAND data improves 
the latter bound substantially~\cite{lisi05}
\be
\sin^2 \theta_{13}  \leq 0.035 ~(0.075), ~ 90\%~  (3\sigma) C.L.. 
\ee 
(This can be compared with  (\ref{limglob}).)

The 1-3 mixing taken at the present upper bound 
modifies the allowed region of the 
parameters $\sin^2\theta_{12} - \Delta m^2_{21}$ obtained from the global 
fit. It shifts the region toward larger $\Delta m^2_{21}$ and smaller 
$\sin^2\theta_{12}$, however the shift  of the best fit point is within 
$1\sigma$~\cite{valen04}. 

It was noticed that the Earth matter regeneration effect has 
strong dependence on $\theta_{13}$,  as $P 
\propto \cos^6\theta_{13} $~\cite{BOS}.
This can be used to determine $\theta_{13}$ in future solar neutrino 
experiments~\cite{BOS,akh}. 
Measurements  of the pp-neutrino fluxes will also improve the 
sensitivity to  $\sin^2\theta_{13}$~\cite{roadmap}. \\

In this paper we study in details possible effects 
of 1-3 mixing on the solar neutrino observables.  
We  analyze   the present determination of 
$\theta_{13}$ and $\theta_{12}$ and 
evaluate the sensitivity of future solar neutrino experiments 
to $\theta_{13}$.

The paper is organized as follows.  
We first (sec.~2) re-derive 
the  $\nu_e-$survival probability including all relevant 
corrections and estimating accuracy of the approximations. 
We consider dependence of the probability 
on $\sin^2\theta_{13}$. In sec.~3 we find simple 
analytic expressions for the probability in the high and low energy limits.   
In sec. 4 we introduce the $\sin^2\theta_{12}- \sin^2\theta_{13}$  
plots and derive the equations for  contours of constant 
values of observables in this plot. 
In sec.~5 we find the allowed region of parameters  $\sin^2\theta_{12}$ 
and  $\sin^2\theta_{13}$ and  consider the bounds 
on these parameters  from individual measurements.  
In sec.~6 the sensitivity of  
future, in particular, the low energy (LowNu)  solar neutrino experiments  
to 1-3 mixing, is estimated. We discuss  possible implications of future 
measurements of $\theta_{13}$ in sec~7.  
Our results are summarized in sec.~8.

%%%%%%%%%%%%%%%%%%%%%%%%%%%%%%%%%%%%%%%%%%%%%%%%%%%%%%%%%%%%%%%%%
\section{The survival probability and $\theta_{13}$}
%%%%%%%%%%%%%%%%%%%%%%%%%%%%%%%%%%%%%%%%%%%%%%%%%%%%%%%%%%%

The  analytic formula  which describes 
conversion of the solar neutrinos in the case of  three neutrino 
mixing and mass split hierarchy,  $\Delta m^2_{31} \gg \Delta m^2_{21}$, 
has been derived long time ago~\cite{3nu,3nuS}. 
Here $\Delta m^2_{31}$  and $\Delta m^2_{21}$ are the mass splits 
associated to  1-3 mixing 
and  to the main channel of the solar neutrino conversion correspondingly.   
The formula was analyzed and further corrected 
recently~\cite{lisi,lim03,BOS}.

In view of future precision measurements  
one needs to have an accurate formula for the probability 
with all relevant corrections included and 
errors due to approximation made estimated. 
For this reason we re-derive the formula 
in a simple and adequate way which employs features of the  
Large Mixing Angle (LMA) MSW solution.   
We will use a very high level of 
adiabaticity~\cite{ms,messiah,parke,smir04} in whole $3\nu$ 
system.\\

Due to  loss of coherence between  the mass eigenstates, 
$\nu_i$ (i = 1, 2, 3), the $\nu_e$ survival probability can be 
written as 
\be
P_{ee} = \sum_{i = 1,2,3} P_{ei}^{sun}  P_{ie}^{Earth}, 
\label{ptot}
\ee
where $P_{ei}^{sun}$ is the probability of $\nu_e \rightarrow \nu_i$ 
conversion in the Sun and 
$P_{ie}^{Earth}$ is the $\nu_i \rightarrow \nu_e$ oscillation 
probability inside the Earth. Strong adiabaticity of the neutrino 
conversion in the Sun leads immediately to 
\be
P_{ei}^{sun} = |U_{ei}^{m0}|^2, 
\label{psun}
\ee
where $U_{ei}^{m0}$ is the $ei$-element of neutrino  mixing matrix 
in matter in the production point. Indeed, $U_{ei}^{m0} \equiv \langle 
\nu_e| \nu_{mi}\rangle$  determines  
the admixture of the eigenstate, $\nu_{mi}$, in  the neutrino 
state in the initial point. 
Then due to  adiabaticity $\nu_{mi}$ transforms to  
$\nu_{i}$ when neutrino propagates to 
the surface of the Sun.  Corrections due to the adiabaticity violation  
are negligible~\cite{smir04}:   
$\Delta P_{ee}/ P_{ee} \sim \gamma^2 \cos 2\theta_{12} /4\sin^2\theta_{12}$,  
where $\gamma$  
%%=  {16 E^2 |\theta_{12}^m|^2}/{\Delta m^2_{21}}$ 
is the adiabaticity parameter.
For $E = 10$ MeV we find $\Delta P_{ee}/ P_{ee} \sim 10^{-8}$.   

Combining (\ref{ptot}) and (\ref{psun}) we obtain  
\be
P_{ee} = \sum_{i = 1,2,3} |U_{ei}^{m0}|^2  P_{ie}^{Earth}. 
\label{ptot1}
\ee
In the absence of the Earth matter effect (the day signal) we have 
\be
P_{ie}^{Earth} = |U_{ei}|^2,
\label{pear}
\ee
where $U_{ei}$ are the elements of the vacuum mixing matrix. 

In what follows we will neglect the Earth matter effect  
on the 1-3 mixing which is determined by 
$2E V_E/ \Delta m^2_{13} < 0.005$, (here $V_E$ is the typical potential 
inside the Earth). Therefore  
\be
P_{3e}^{Earth} \approx |U_{e3}|^2. 
\label{pear2}
\ee
The Earth matter effect on two other probabilities can be described by  
the term $F_{reg}$ defined as the deviation of probability 
$P_{2e}^{Earth}$ from the no-oscillation one:  
\be
P_{2e}^{Earth} = |U_{e2}|^2 + F_{reg}. 
\label{pear3}
\ee
Then the unitarity condition, $\sum_{i = 1,2,3} P_{ie}^{Earth} = 1$ 
and (\ref{pear2}) lead to  
\be
P_{1e}^{Earth} = |U_{e1}|^2 - F_{reg}.
\label{pear1}
\ee
Plugging  (\ref{pear}, \ref{pear2}, \ref{pear1}) into 
(\ref{ptot1}) we obtain 
\be
P_{ee} = \sum_{i = 1,2,3} |U_{ei}^{m0}|^2  |U_{ei}|^2
+ F_{reg} \left(|U_{e2}^{m0}|^2 - |U_{e1}^{m0}|^2\right).  
\label{ptot2}
\ee

We will use the standard parameterization of the 
mixing matrix in terms of mixing angles:  
\be
|U_{e1}| \equiv \cos \theta_{13} \cos \theta_{12},~~~~ 
|U_{e2}| \equiv \cos \theta_{13} \sin \theta_{12},~~~~ 
|U_{e3}| \equiv \sin \theta_{13}, 
\label{mixpar}
\ee
and similar definition holds for the matrix elements in matter with 
substitution $\theta_{13} \rightarrow \theta_{13}^{m0}$ and 
$\theta_{12} \rightarrow \theta_{12}^{m0}$.  
The angle $\theta_{12}$ is identified  with the ``solar mixing angle''. 

In terms of mixing angles the probability (\ref{ptot2}) can be rewritten 
as 
\be
P_{ee} = \cos^2\theta_{13}^{m0} \cos^2\theta_{13} P_{ad} + 
\sin^2\theta_{13}^{m0}\sin^2\theta_{13} - 
F_{reg}\cos^2\theta_{13}^{m0} \cos 2\theta_{12}^{m0}, 
\label{pro}
\ee
where $P_{ad}$ is the usual two neutrino adiabatic probability: 
\be 
P_{ad} \equiv \sin^2\theta_{12} + 
\cos^2 \theta_{12}^{m0} \cos 2\theta_{12}. 
\ee 
For $F_{reg} = 0$ the probability (\ref{pro}) coincides with the one obtained  
in \cite{3nuS}. \\

Let us find the mixing angle $\theta_{13}^{m0}$ in matter.  
The Hamiltonian of 3$\nu$ system 
can be written in the flavor basis as
\be
H = U_{23} U_{13} U_{12} H^d U_{12}^{\dagger}U_{13}^{\dagger}
U_{23}^{\dagger} + V_3. 
\ee
Here $H^d = diag(0,~ \Delta m_{21}^2/2E,~  \Delta m_{13}^2/2E)$, 
 $V_3 \equiv  diag(V, 0, 0)$, and    
$V \equiv \sqrt{2} G_F \rho Y_e$, where $\rho$ is the matter density and 
$Y_e$ is the electron number density fraction.

Performing rotations $U_{23} U_{13}$ 
we arrive at the basis $(\nu_{e}', \nu_{2}', \nu_{3}')$ in which the
Hamiltonian   becomes
\be
H = U_{12} H^d U_{12}^{\dagger} + U_{13}^{\dagger} V U_{13}. 
\label{ham1}
\ee
If the matter effect is small we can neglect mixing of  the third
state (off-diagonal terms of the matrix $U_{13}^{\dagger} V U_{13}$). 
This state then decouples and for the rest of the system the
Hamiltonian becomes
 \be
H_2 = U_{12} H^d U_{12}^{\dagger} + \cos^2 \theta_{13} V_2, 
\label{2mix}
\ee
where $V_2 \equiv diag(V, 0)$.

In the lowest approximation the matter effect on the 1-3 mixing
can be found making  an additional 1-3 rotation 
which eliminates the 1-3 elements in (\ref{ham1}).  
The angle of rotation, $\theta_{13}'$, is given by
\be
\tan 2 \theta_{13}' = \frac{V \sin 2\theta_{13}}{\Delta m^2_{13}/(2E) - V
\cos 2\theta_{13} + \Delta m^2_{21}/(2E) \sin^2 \theta_{12}} \approx 
\sin 2\theta_{13}\epsilon_{13},  
\label{thet}
\ee
where 
\be 
\epsilon_{13} \equiv  \frac{2EV}{\Delta m^2_{13}} = 
0.062  \left(\frac{E}{10 {\rm MeV}}\right)
\left(\frac{\rho Y_e}{100 {\rm g/cc} } \right) 
\left(\frac{2.5 \cdot 10^{-3}{\rm eV^2}}{\Delta m^2_{13}}\right).  
\label{eps}
\ee
So, approximately, $\theta_{13}' = \theta_{13} \epsilon_{13}$. 

The additional 1-3 rotation generates negligible 
2-3 elements~\footnote{Elimination of this term requires an additional 2-3 
rotation by  the angle 
$\theta_{23}' = 2E H_{23}/\Delta m^2_{31} < 10^{-4}$.} 
\be
H_{23} = \frac{V}{4} \sin 2\theta_{13} \sin 2\theta_{12} 
\frac{\Delta m^2_{21}}{\Delta m^2_{13}}.  
\label{next}
\ee
It also   
leads to corrections to the 1-2 block  of the Hamiltonian: 
\be 
H_2 = U_{12} H^d U_{12}^{\dagger} + 
\cos^2 \theta_{13} \left(1 -  
\sin^2 \theta_{13} \epsilon_{13}\right)V_2. 
\label{2ham}
\ee

In the first order in $2VE/\Delta m_{13}^2$
the rotation $\theta_{13}'$ decouples the third state.   
So, the 1-3 mixing angle in matter is given by
\be
\theta_{13}^m  = \theta_{13} + \theta_{13}' + 
O(\theta_{13}'\epsilon_{13}), 
\ee
and  using (\ref{thet}) we obtain 
\be
\sin^2 \theta_{13}^m  \approx  \sin^2 \theta_{13} 
\left(1 + 2 \epsilon_{13} \right) + 
O\left(\sin^2 \theta_{13} \epsilon_{13}^2,~
\sin^4 \theta_{13}\epsilon_{13} \right).   
\label{13corr}
\ee
%%%%%%%%%%%%%%%%%%%%%%
(See also \cite{lisi}.) For typical energy $E = 10$ MeV 
the relative matter correction is (10 - 15)\%.   
The correction is negligible at low energies. 
%%%%%%%%%%%%%%%%%%%%%%%

The 1-2 mixing in matter is determined from diagonalization of 
(\ref{2mix}) or (\ref{2ham}).  Neglecting very small correction 
($\sim  \epsilon_{13}\sin^2\theta_{13}$) in (\ref{2ham})  
one obtains for $\theta_{12}^m$ the standard  
$2\nu$ formula with substitution 
$V \rightarrow V\cos^2 \theta_{13}$.\\   

Let us find  dependence of  $F_{reg}$ on the 1-3 mixing explicitly. 
The probability of $\nu_2 \rightarrow \nu_e$ conversion 
can be written in terms of the amplitudes of transitions 
between the mass states  $\nu_2 \rightarrow \nu_i$  , $A_{i2}$,  as 
\be
P_{2e} = |A_{12} U_{e1} + A_{22}  U_{e2}|^2 = 
\cos^2 \theta_{13}  |A_{12} \cos \theta_{12} + A_{22} \sin 
\theta_{12}|^2. 
\ee
Since in the absence of  matter effect $A_{12} = 0$ and  $A_{22} = 1$,  
in the lowest approximation in $A_{12} \propto V$,  we obtain   
\be
F_{reg} = \cos^2 \theta_{13} \sin 2 \theta_{12} Re A_{12}. 
\label{Freg}
\ee
The transition  amplitude $A_{12}$ should be found by solving  the 
evolution equation with the  Hamiltonian (\ref{2ham}).   
It   is proportional to the potential:  
$A_{12} \propto  V \cos^2\theta_{13}$, and consequently,   
 $F_{reg}$ (\ref{Freg}) can be rewritten as 
\be
F_{reg} = \cos^4 \theta_{13} f_{reg} + O(f_{reg}^2), 
\label{Freg2}
\ee
where 
\be
f_{reg} \equiv \sin 2 \theta_{12} Re A_{12}'
\ee
is the  $\nu_e-$regeneration factor  
given by the  $2\nu$ amplitude $A_{12}' 
\equiv A_{12}/ \cos^2 \theta_{13}$,  
and it does not depend on $\theta_{13}$.

The regeneration factor integrated with the exposure function 
over the zenith angle can be estimated as  
\be
\bar{f}_{reg} \sim \frac{E V_0 \sin^2 2\theta_{12}}{4 \Delta m^2_{21}},   
%\sin \Phi_0 \sum_{i = 0}^{i = n-1} \Delta V_i \sin\Phi_i \sim 1\%. 
\ee
where $V_0$ is the potential at the surface of the Earth. 
Indeed, the integrated effect is mainly due to the 
oscillations in the mantle of the Earth along the trajectories 
which are not too close to the horizontal ones. 
(For the core crossing trajectories the effect of the core is attenuated 
\cite{araic}.) In turn, for the mantle trajectories the adiabaticity is 
nearly fulfilled~\cite{smir04} and therefore characteristics of 
oscillations 
(averaged probability and depth of oscillations) are determined by the 
potentials in the initial and final points of the trajectory, 
that is, by the potential near the surface of the Earth 
(for numerical consideration see~\cite{GPS}).

Inserting the expression for $F_{reg}$ into (\ref{pro}) we obtain 
\be
P_{ee} = \cos^2\theta_{13}^{m0} \cos^2\theta_{13} 
(P_{ad} - \cos^2 \theta_{13} \cos 2\theta_{12}^{m0} f_{reg})  +
\sin^2\theta_{13}^{m0} \sin^2\theta_{13}.  
\label{pro1}
\ee
Taking the mixing angle $\theta_{13}^m$ in the first order 
in $\epsilon_{13}$ we find 
\be
P_{ee} \approx \cos^4\theta_{13}(1 - 2\tan^2\theta_{13}\epsilon_{13})
(P_{ad} - \cos^2 \theta_{13} \cos 2\theta_{12}^{m0} f_{reg})  +
\sin^4\theta_{13}~. 
\label{pro2}
\ee
Notice that the pre-factor 
$(1 - 2\tan^2\theta_{13}\epsilon_{13}(E))$ which describes the  matter effect on 
the 1-3 mixing differs in the high and low energy limits. \\

The expression~(\ref{pro1}) can be rewritten as 
\be
P_{ee}  \cong \cos^2 \theta_{13}^m \cos^2 \theta_{13} P_{ee}^{(2)} 
(\Delta m^2, \theta_{12}, \theta_{13})
 + \sin^2\theta_{13}^m \sin^2\theta_{13}, 
\label{3genpee}
\ee
where 
$$
P_{ee}^{(2)} =  P_{ad} - \cos^2 \theta_{13} \cos 2\theta_{12}^{m0} f_{reg}
$$
is the total $\nu_e$ survival probability for
$2\nu$ mixing which includes both the conversion in the Sun and 
the oscillations in  the Earth.  
$P_{ad}$ depends on $\theta_{13}$ via the effective matter potential 
$V \cos^2\theta_{13}$.\\  

The probability should be averaged over the distribution of the neutrino 
sources in the Sun. 
For the $K$-component of the solar neutrino spectrum 
($K \equiv pp, Be, pep, B, N, O$)  the averaged  probability   
can be written as~\cite{smir04}  
\be 
P_{ee,K}^{(2)} = \sin^2 \theta + \cos^2\theta_m (V_K) \cos 2\theta - 
0.5 \delta_K \cos 2\theta_m(V_K) \cos 2\theta - \cos 2\theta_m(V_K) 
\cos^2 \theta_{13} f_{reg},   
\label{surv2}
\ee 
where $V_K$ is the effective value of the matter potential 
in the region of production of the $K$  neutrino component. 
$\delta_K$ is an additional correction due to the 
integration over the production region~\cite{smir04},
and in what follows we will neglect it.\\ 

Notice there are  three features of dynamics of propagation 
which allow one  
to immediately derive the $3\nu$ survival probability in terms of
the $2\nu$ probabilities:

1). Adiabaticity of  propagation of the third neutrino; 

2). Negligible Earth matter effect on 1-3 mixing;  

3). Averaging of oscillations and loss of coherence
of the third mass eigenstate.

\noindent
Essentially this means  
that the third neutrino eigenstate decouples from dynamics and
propagates independently. Its contribution to the probability adds
incoherently.
Therefore the problem is reduced to two neutrino
problem and additional factors in (\ref{pro1}) 
are just projection of the flavor  
basis onto the basis of states which 
contains the decoupling state $\nu_{3m}$.
In (\ref{3genpee}) factors  $\sin^2 \theta_{13}^{m0}$ and 
$\cos^2 \theta_{13}^{m0}$ are from the projection in the initial state, 
whereas the factors  $\sin^2 \theta_{13}$ and 
$\cos^2 \theta_{13}$ are from projection in the final state.\\

Up to negligible corrections the expression for the 
probability (\ref{3genpee}) can be rewritten in form  
\be
P_{ee}  \cong (1 - \sin^2\tilde{\theta}_{13}^2)^2  P_{ee}^{(2)}
(\Delta m^2, \theta_{12}, \tilde{\theta}_{13})
 + \sin^4 \tilde{\theta}_{13},
\label{3genpee1}
\ee
where 
\be
\sin^2 \tilde{\theta}_{13} \equiv \sin^2 \theta_{13} (1 + 
\epsilon_{13}), 
\label{sin1}
\ee
which coincides with the form without matter corrections to 
$\theta_{13}$~\footnote{Strictly, this is correct in the high energy 
or  low energy limits, where $P_{ee}^{(2)} \sim  P_{ad}$ 
depends very weakly on the potential. At low energies the  
matter corrections are negligible anyway (see eq. (\ref{eps}).}.  
This means that the matter effect on the 1-3 mixing in the 
probability can be accounted by renormalization of $\sin^2\theta_{13}$. 
Inversely, using usual formula  for the probability without 
matter corrections in the analysis of the data we determine  
$\tilde{\theta}_{13}$ averaged over the relevant energy interval. Then 
the true vacuum mixing angle  will be smaller:  
\be
\sin^2\theta_{13} = \sin^2 \tilde{\theta}_{13}(1 - \bar{\epsilon}_{13}).
\label{sin2}
\ee
Here $\bar{\epsilon}_{13} = \langle \epsilon_{13} \rangle_E$ 
is the value averaged over the  relevant energy range. 
Renormalization depends on the neutrino energy:  
$\bar{\epsilon}_{13} \sim 0.05$ at high 
energies and $\bar{\epsilon}_{13} \sim 0.002$ in the $pp$-neutrino range, 
and the latter can be neglected.

\section{Two limits}
%%%%%%%%%%%%%%%%%%%%%%%%%%%%%%%%%%%%%%%%%%%%%%%%%%%%%%%%%%%%%%%%%%%

As we will see,  the expected sensitivity of solar neutrino 
studies to the 1-3 mixing will not be better than 
$\sigma (\sin^2\theta_{13}) \sim  0.01$. 
According to (\ref{3genpee}) that would correspond to the  change of the 
survival probability 
\be
\delta P \approx - 2 P \sigma(\sin^2 \theta_{13}) \sim 1\%.  
\ee
Therefore in the present discussions we can neglect 
various corrections to the 
probability which are much smaller that 0.01. 
In particular, 

\begin{itemize} 

\item
the term $\sin^4\theta_{13} < 0.002$ can 
be neglected  in eq. (\ref{3genpee});

\item 
the matter effect on the 1-3 mixing is small and can be taken into 
account as a small renormalization of $\sin^2 \theta_{13}$ (see sec. 2); 

\item 
we will neglect $f_{reg}^2 < 10^{-3}$ corrections in $F_{reg}$. 

\end{itemize}

In qualitative consideration we will  use 
simplified expressions for the survival probability obtained in the limits 
of  high and low energies.  
In fact,  all relevant fluxes are either in the low  
(pp, Be, N, O, pep) or in the high,  
$E > 5$ MeV (B, hep),  energy limits.  
The expansion parameter equals  
\be
\eta \equiv \frac{\Delta m^2_{21}}{2 E V}  
\ee
at  high energies  and  $1/\eta$ at  
low energies~\footnote{Notice that in \cite{smir04} we use 
different definition  of $\eta$: $\eta \rightarrow 1/\eta'$.}.  
\\

1). {\it In the high energy limit},  performing expansion of 
(\ref{surv2}) in $\eta$ we find the survival probability 
\be
P_{ee}^{(h)} 
\approx \cos^4 \theta_{13} \sin^2\theta_{12}   
+ \frac{1}{4} \cos 2\theta_{12} 
\sin^2 2\theta_{12} \eta^2 +  \cos^6 \theta_{13} f_{reg}.  
\label{highl}
\ee
Here the first term is due to the non-oscillatory adiabatic conversion.
In the second term,  which describes the effect of averaged oscillations,   
dependence on $\theta_{13}$ is absent. 
In contrast,  the third (regeneration) term has the strongest dependence 
on  $\theta_{13}$~\cite{BOS}. It  originates from the 
common factor $\cos^4 \theta_{13}$ and the effective potential 
in the $3\nu$ case which contains another   $\cos^2 \theta_{13}$ 
factor. 

In the lowest order in $\sin^2\theta_{13}$ we can rewrite (\ref{highl}) 
as 
\be
P_{ee}^{(h)}
\approx \sin^2\theta_{12}
+ \frac{1}{4} \cos 2\theta_{12}
\sin^2 2\theta_{12}\eta^2  + f_{reg} - \sin^2 \theta_{13}(2\sin^2\theta_{12} 
+3f_{reg}). 
\label{highl2}
\ee
Neglecting the oscillation corrections  (second term) and the double 
suppressed term $\sim \sin^2 \theta_{13} f_{reg}$ we obtain  
\be
P_{ee}^{(h)}
\approx  (1 - 2\sin^2\theta_{13})\sin^2\theta_{12} + f_{reg}. 
\label{highl3}
\ee \\

2). {\it For the low energy neutrinos} ($E < 2$ MeV),   
the matter effect is small and 
in the  lowest order in  $\eta^{-1}$  we have 
\be
P_{ee}^{(l)} \approx 
\cos^4 \theta_{13} (1 - 0.5 \sin^2 2\theta_{12}) - 
0.5 \cos^6 \theta_{13} \cos 2\theta_{12}\sin^2 
2\theta_{12} \eta^{-1}. 
\label{lowl}
\ee
Here the first term is the effect of averaged vacuum oscillations.  
The second one gives the matter effect correction which contains strong 
dependence on $\theta_{13}$. An additional second power of 
$\cos \theta_{13}$ originates from the matter potential. 
The regeneration effect is neglected.  
Numerical calculations confirm that the linear in 
$\eta^{-1} \propto E$  approximation 
(\ref{lowl}) works well up to $2$ MeV.

In the lowest order in $\sin^2\theta_{13}$ we find from (\ref{lowl})
\begin{eqnarray}
P_{ee}^{(l)} \approx
1 - 0.5 \sin^2 2\theta_{12} - 0.5 
\cos 2\theta_{12}\sin^2 2\theta_{12}\eta^{-1} - \\
- \sin^2 \theta_{13} \left[2 - \sin^2 2\theta_{12} -
1.5 \cos 2\theta_{12}\sin^2 2\theta_{12}\eta^{-1}
\right]. 
\label{lowl2}
\end{eqnarray}
For very low energies (pp-neutrino  range) the matter corrections can be 
neglected, so that  
\be
P_{ee}^{(l)} \cong (1 - 2 \sin^2 \theta_{13}) (1 - 0.5\sin^2 
2\theta_{12}). 
\label{3genpeepp}
\ee

From  Eqs. (\ref{highl} - \ref{lowl2}) we conclude the following.    

\begin{itemize}

\item
The dependences of the  probability on $\theta_{12}$ 
for the high energies and
low energies are opposite: $P_{ee}^{(l)}$ increases whereas 
$P_{ee}^{(h)}$ decreases with decrease of $\theta_{12}$.    

\item 
In contrast, $P_{ee}^{(l)}$ and $P_{ee}^{(h)}$ both decrease with 
increase of  $\theta_{13}$. So,  $\theta_{12}$ and $\theta_{13}$ 
correlate at low energies and anti-correlate at high energies. 
For the $^{8}{B}$ neutrinos an increase of  $\theta_{13}$ 
can be compensated by increase of $\theta_{12}$, whereas 
at low energies (for the $pp$ neutrinos), 
 $\theta_{12}$ should decrease to compensate increase of $\theta_{13}$. 

\item 
Both for the high and low neutrino energies  
the dependence of  probability  
on $\Delta m^2_{21}$ appears via  small corrections only, 
and therefore, is weak.   

\end{itemize}

Apparently strong dependence on $\theta_{13}$ cancels  in 
the ratio of probabilities (or the ratio of the corresponding 
observables): 
\be
\frac{P_{ee}^{(h)}}{P_{ee}^{(l)}} = 
\frac{\sin^2\theta_{12} 
+ 0.25 \cos^{-4} \theta_{13} 
\cos 2\theta_{12} \sin^2 2\theta_{12}\eta^2 + 
\cos^2 \theta_{13} f_{reg}}{1 - 0.5 \sin^2 2\theta_{12}}. 
\label{rprob}
\ee
Notice that  dependence of the ratio on $\theta_{13}$ appears via   
small corrections and it can be neglected in the first approximation. 
So, measurements of the ratio  gives $\sin^2\theta_{12}$. 
Then one of the probabilities (in low or high limits) can be used to 
find $\sin^2\theta_{13}$.

%%%%%%%%%%%%%%%%%%%%%%%%%%%%%%%%%%%%%%%%%%%%%%%%%%%%%%%%%%%%%%%
\section{$\sin^2\theta_{12} - \sin^2\theta_{13}$ plots 
and iso-contours of observables }
%%%%%%%%%%%%%%%%%%%%%%%%%%%%%%%%%%%%%%%%%%%%%%%%%%%%%%%%%%%

There are two important features of the problem: 

\begin{itemize}

\item
Weak dependence of the probability and observables on 
$\Delta m_{21}^2$ within the allowed region; 
%Notice that already present data determines $\Delta m^2_{21}$ 
%with rather high accuracy. 

\item
$\sin^2\theta_{12} - \sin^2\theta_{13}$ degeneracy. 

\end{itemize}

Let us consider these features in some  details. \\ 

1). The \kl spectral data has shown  high precision in 
determination of $\Delta m^2_{21}$.  The  relative error 
\be
\delta_{\Delta} \equiv \frac{\delta(\Delta m^2_{21})}{\Delta m^2_{21}} 
\ee
is about $\delta_{\Delta} = 12\%$ (3$\sigma$)   
at the present, and with 3 kTy statistics 
from \kl this can further reduce down to 7\% \cite{th12new}. 

As follows from (\ref{highl}) and (\ref{lowl2}),  
variation of the probability in the high energy limit with 
$\Delta m^2_{21}$ equals  
\be
\delta P^{(h)}_{ee} = \left[ \frac{1}{2} \eta^2 \cos 2 \theta_{12} \sin^2 
2\theta_{12} 
- \cos^6 \theta_{13} f_{reg}\right] 
\delta_{\Delta}.   
\label{vardmh}
\ee
For  $E = 10$ MeV  and  
$\delta_\Delta \sim 0.1$,   
we find $\delta P \sim 0.005$.\\ 

In the low energy part,  variations of the probability 
with $\Delta m^2_{21}$ are even  weaker:   
\be
\delta P_{ee}^{(l)} \approx
\frac{1}{2} \eta^{-1}\cos^6 \theta_{13}   
%\left(\frac{E V}{\Delta m^2_{21}}\right)
\cos 2\theta_{12}\sin^2 2\theta_{12} \delta_{\Delta}. 
%\frac{\delta(\Delta m^2_{12})}{\Delta m^2_{21}}. 
\label{varl2}
\ee
For $E = 0.4$ MeV we find 
$\delta P_{ee}^{(3)} = 0.011 \delta_{\Delta} \approx 0.001$. \\

2). According to  the discussion in sec. 2,  
in the first approximation, all  
experiments which probe the high energy part of the spectrum 
are sensitive to the same combination of 
mixing angles $\theta_{12}$ and $\theta_{13}$. Similar  
statement is true for the low energy experiments 
but  the combination of mixing angles is different.  
Therefore to determine $\sin^2\theta_{12}$ and $\sin^2\theta_{13}$ separately 
one needs to employ  both types of 
experiments. \\
 
In view of these two features,  we will  present results of
our study in $\sin^2\theta_{12} - \sin^2\theta_{13}$ plane
for fixed values of $\Delta m^2_{21}$.\\

Let us  construct the contours of constant 
values of various observables in the 
$\sin^2\theta_{12} - \sin^2\theta_{13}$ plane. 
Essentially,  the iso-contours  for a given observable $X$ coincide with 
the 
contours of constant survival probability averaged over the relevant 
energy range with the corresponding cross-section: 
\be
\langle R_{ee} \rangle_X = C_X. 
\ee
The survival probabilities found in the high and low energy limits 
determine two different types of the iso-contours.   
%($h-$ and $l-$).  
For the high energy observables ($E > 5$ MeV), using Eq. 
(\ref{highl2}), we find an analytic expression for the contours of 
constant  
$X$: 
\be
\sin^2 \theta_{13} \approx 
\frac{\sin^2\theta_{12} +  0.25 \cos 2\theta_{12}\sin^2 
2\theta_{12} \eta^2_{X} + 
\langle f_{reg} \rangle_{X} - C_{X}}{
2\sin^2\theta_{12} + 3 \langle f_{reg} \rangle_{X}}, 
\label{con-h}
\ee
where 
\be 
\eta^2_{X} \equiv \left\langle \left(\frac{\Delta m^2}{2E V_B} \right)^2 
\right\rangle_X
\ee
is the averaged over the neutrino energy value of $\eta^2$ 
folded with the cross-section,   
energy resolution of  detector,  {\it etc.}.  
In the allowed range of $\sin^2\theta_{12}$ (with the average value of 
mixing denoted by $\bar{\theta}_{12}$) the strongest dependence 
comes from  the denominator of (\ref{con-h}):
\be
\sin^2 \theta_{13} \sim  A (\sin^2\theta_{12} - C_{X}),~~~ 
A = (2 \sin^2\bar{\theta}_{12} + 3 {f}{reg})^{-1}.
\ee 
So, in the first approximation the contours are  straight lines.  
This reproduces well the results of numerical 
calculations presented in sec. 5, 6. Notice that $\sin^2 \theta_{13}$ 
increases with $\sin^2\theta_{12}$.\\ 

For the low energy observables, $Y$, using (\ref{lowl2}), we obtain 
expressions for the  iso-contours as 
\be
\sin^2 \theta_{13} \approx 
\frac{1 - 0.5 \sin^2 2\theta_{12} - 0.5 \cos 2\theta_{12} \sin^2 
2\theta_{12}\eta_{Y}^{-1} - 
C_{Y}}{2  - \sin^2 2\theta_{12} - 1.5  \cos 2\theta_{12}\sin^2 
2\theta_{12}\eta_{Y}^{-1}}, 
\label{con-l}
\ee
where 
\be
\eta_{Y}^{-1} \equiv \left\langle \frac{2 E V}{\Delta m^2_{12}} 
\right\rangle_{Y}.
\ee 
In the rough approximation  
\be
\sin^2 \theta_{13}\propto  1 - C_Y - 2\sin^2\theta_{12}(1 - 
\sin^2\theta_{12})   
\ee
and it has minimum at $\sin^2\theta_{12} = 0.5$.  Other terms in 
(\ref{con-l}) shift this minimum  
to smaller values of $\sin^2\theta_{12}$. So, in the region of interest, 
$\sin^2\theta_{12} \sim  0.3$,  the lines of constant values of 
observables 
have negative slope: $\sin^2\theta_{13}$ decreases with $\sin^2\theta_{12}$. 
%The equation  (\ref{con-l}) reproduces well the  iso-contours 
%presented in the next section. 

%%%%%%%%%%%%%%%%%%%%%%%%%%%%%%%%%%%%%%%%%%%%%%%%%%%%%%%%%%%%%%%%%%%%%%%%%%%%%%
\section{Constraints on the mixing parameters from the present experiments}
%%%%%%%%%%%%%%%%%%%%%%%%%%%%%%%%%%%%%%%%%%%%%%%%%%%%%%%%%%%%%%%%%%%%%%%%%%%%%

%%%%%%%%ffff1%%%%%%%%%%%%%%%%%%%%%%%%%%%%%%%%%%%%%%%%%%%%%%%%%%%%%%%%%%%
\begin{figure}
\includegraphics[width=12.0cm, height=14cm]{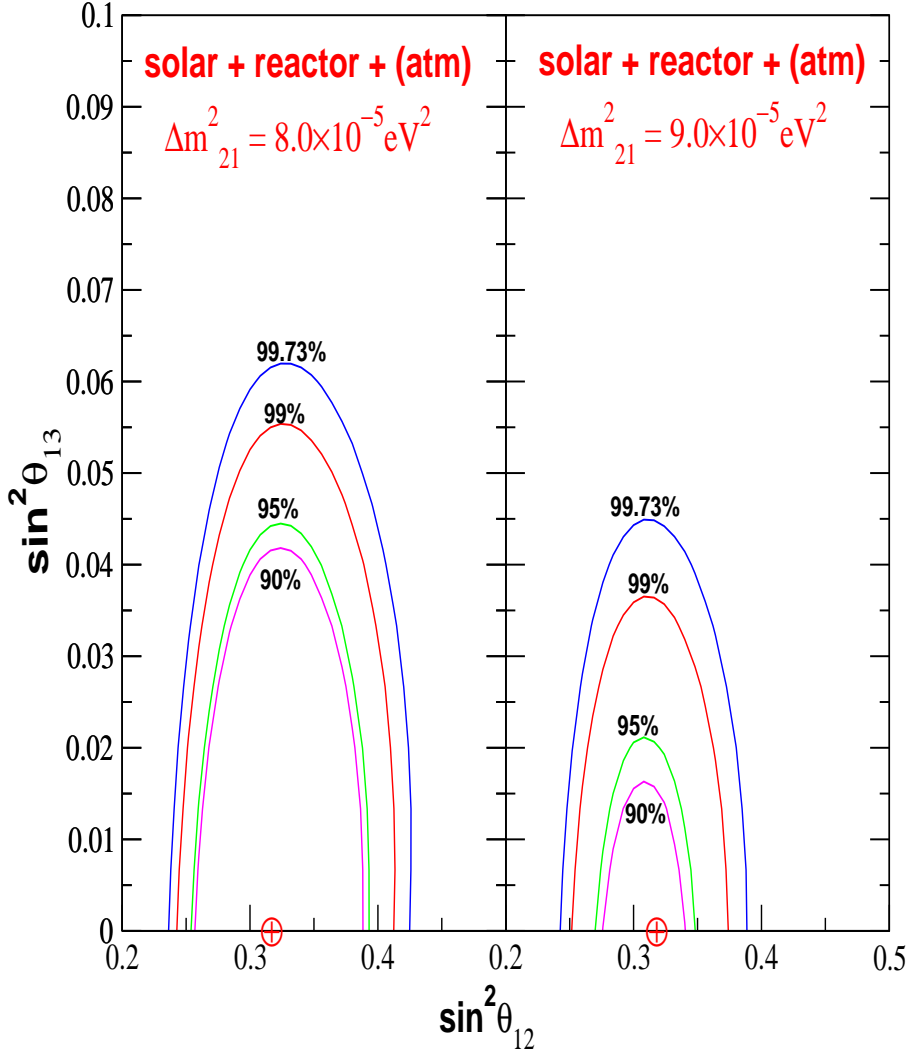}
\caption{The 90\%, 95\%, 99\% and 99.73\% C.L. 
allowed regions of mixing angles
from the global $\chi^2$ analysis of the solar neutrino,
\kl and CHOOZ  neutrino
data.
The left panel is  for $\dm =8 \cdot 10^{-5}$ eV$^2$
and the right panel is for $\dm = 9 \cdot 10^{-5}$ eV$^2$.
The global best-fit point obtained by letting all the parameters (including 
$\Delta m^2_{21}$) vary freely is marked by the crossed circle.}
\label{contourglob_s12s13}
\end{figure}
%%%%%%%%%%%%%%%%%%%%%%%%%%%%%%%%%%%%%%%%%%%%%%%%%%%%%%%%%%%%%%%%

%%%%%%%%%%%%%%%%%%%%%%%%%% ffff2%%%%%%%%%%%%%%%%%%%%%%%%%%%%%%%%%%%%%%%%
\begin{figure}
\includegraphics[width=12.0cm, height=14cm]{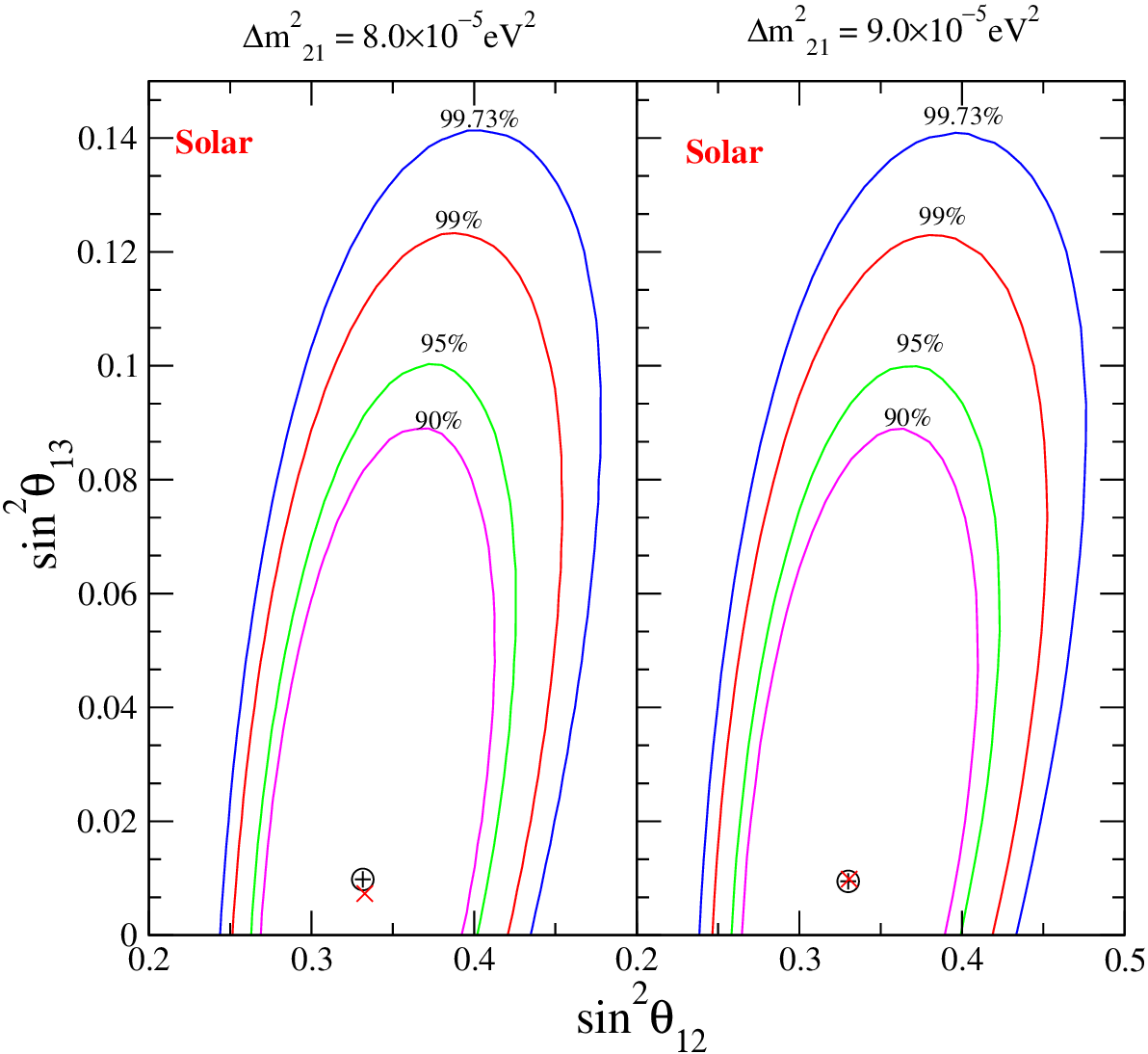}
\caption{The 90\%, 95\%, 99\% and 99.73\% C.L. 
allowed regions of mixing angles
from the global $\chi^2$ analysis of the solar neutrino data only.
The left panel is  for $\dm =8 \cdot 10^{-5}$ eV$^2$
and the right panel is for $\dm = 9 \cdot 10^{-5}$ eV$^2$.
The global best-fit point obtained by letting all the parameters vary freely
is marked by $\oplus$. Also shown by crosses are the local best-fit points
obtained at fixed $\dm$. }
\label{contoursol_s12s13}
\end{figure}
%%%%%%%%%%%%%%%%%%%%%%%%%%%%%%%%%%%%%%%%%%%%%%

In this section we describe the results of our numerical computations. 

We first perform the global analysis of different data samples. 
Some details of the analysis follow. 
The CHOOZ $\chi^2$ depends on $\Delta m^2_{31}$ and  
in our fit we vary this parameter freely in the 
range allowed by the zenith angle SuperKamiokande atmospheric 
neutrino data~\cite{kearns}. 
The details of procedure for the solar and CHOOZ
data can be found in \cite{kkthree} while the \kl analysis 
follows the method outlined  in 
\cite{kl162us,kl766us}. We use the  \kl data from the latest version 
of paper~\cite{v3} where new source of the background has been 
taken into account.  

For the solar neutrinos we include the data on total rates from the 
radiochemical 
experiments Cl \cite{cl} and Ga (SAGE, GALLEX and GNO combined)
~\cite{ga}, 
the Superkamiokande day-night spectrum data \cite{sksolar},  
the SNO CC (charged current), ES (neutrino-electron scattering) and NC 
(neutral current) rates from the salt phase-I, the combined CC+ES+NC 
energy spectrum from the pure D$_2$O phase \cite{sno} as well as the day-night 
spectrum data of CC,ES and NC events from the salt-phase II \cite{salt2}. 

The $^8B$ neutrino flux normalization,  $f_B \equiv F_B /F_B^{SSM}$, 
where  $F_B$ is the true flux  and $F_B^{SSM}$ is the flux 
predicted in the SSM~\cite{bp04}, is left as a free parameter in the fit.  
Essentially $f_B$ is fixed  by the NC data from the 
SNO experiment. 
For the other fluxes and their uncertainties we use the 
predictions from the Standard Solar Model (SSM)~\cite{bp04}.  

In fig. \ref{contourglob_s12s13} 
we show the allowed regions of the parameters $\sin^2\theta_{12} - 
\sin^2\theta_{13}$ (for two different 
values of $\Delta m_{12}^2$) which follow from the global fit of 
all available  
solar neutrino data as well as data from reactor experiments KamLAND 
and CHOOZ. The contours are marginalised over $f_B$.    
The global minimum has been  obtained by 
letting all the parameters (including \dm) vary freely in the fit.  
With this,  the values of parameters in  the global minimum
come at
\be
\dm = 8.0 \cdot 10^{-5}~~{\mathrm {eV^2}}, ~~
\sin^2\theta_{12} = 0.31, ~~\sin^2\theta_{13} = 0.00, ~~f_B=0.84. 
\label{globm}
\ee

The two panels in fig. \ref{contourglob_s12s13}
correspond to two fixed illustrative values of $\Delta m^2_{21}$ 
from the current allowed range. 
The contours are plotted with respect to the global minimum 
(\ref{globm}) using the 
definition of $\Delta \chi^2$ which corresponds to 3 parameters.
The upper bound,  $\sin^2\theta_{13} < 0.055$ (99\% C.L.)  
is tighter than the one obtained without including the phase-II 
SNO salt data (0.061).  
%The later is in agreement with results of other analyses~\cite{valen04,john}. 
%}}
%%
According to fig. \ref{contourglob_s12s13} with increase of $\Delta m^2_{21}$ 
the allowed region shifts to smaller $\sin^2\theta_{12}$;  
with increase of $\sin^2\theta_{13}$  
the value of $\sin^2\theta_{12}$ in minimum of $\chi^2$ practically does 
not change.\\    

We have performed the global fit of the solar neutrino data only.
The best-fit values of parameters are found to be 
\be
\Delta m^2_{21} = 6.4 \times 10^{-5} {\mathrm {eV^2}},~~~~~
\sin^2\theta_{12} = 0.33, ~~~~~~ \sin^2\theta_{13} = 0.01, ~~~~~
f_B = 0.84.   
\label{globs}
\ee
Now the 1-3 mixing is non-zero but statistically insignificant.  
%%
%This is in agreement with $\sin^2\theta_{13} = 0.02$ obtained in 
%\cite{valen04}. 
The solar $\chi^2$ is somewhat flat in the region 
$\sin^2\theta_{13} = 0 - 0.03$.   
For $\sin^2\theta_{13} = 0.01$ and 0.02 we obtain the $\chi^2_{min}$ =   
113.84 and 114.04 correspondingly. 

Notice that we find small boron neutrino flux: 
$f_B = 0.84$ in the units of flux predicted in~\cite{bp04}. 
Recent calculations of fluxes with new radiative opacities and 
the heavy element abundances give even smaller value $f_B = 0.78$ 
\cite{bahcall05}.

In fig.~\ref{contoursol_s12s13} we show the constant $\chi^2$ contours 
obtained from global analysis of 
the solar neutrino data only and when $\Delta m^2_{21}$ is fixed. 
The contours have been calculated with respect to the global minimum 
(\ref{globs}) and using the $\Delta \chi^2$ 
values corresponding to 3 parameters. We also show 
the local minima found for 
fixed values of $\Delta m^2_{21}$. In particular, for 
$\Delta m^2_{21} = 8\cdot 10^{-5}$ eV$^{2}$ we obtain 
\be
\sin^2\theta_{13} = 0.007~^{+ 0.080}_{-0.007} ~~~(90 \% ~{\rm C.L.}). 
\label{bfsol}
\ee

The best fit value of  $\theta_{13}$ changes with $\Delta m^2_{21}$ very  
weakly,  e.g., for $\Delta m^2_{21} = 9 \cdot 10^{-5}$ eV$^{2}$ we find 
$\sin^2\theta_{13} = 0.01$.  
The fig. \ref{contoursol_s12s13} shows also that at present 
the solar data alone has a weak sensitivity to $\sin^2\theta_{13}$: 
the 90\% C.L. bound equals  $\sin^2\theta_{13} \leq 0.087$. 
Inversely, the values of $\sin^2\theta_{13}$ allowed by the 
global fit (\ref{limm}) have weak impact on the global analysis of the 
solar neutrino data. From fig. \ref{contoursol_s12s13} we find that 
with increase of $\sin^2\theta_{13}$ from 
0 to 0.05 the best fit value and $90\%$ allowed region of 
$\sin^2\theta_{12}$ shift to larger values by 
$\Delta \sin^2\theta_{12} = 0.02$ only which is smaller than 
$1\sigma$ error and in agreement with earlier estimations in 
\cite{concha}. \\

Let us now consider bounds on mixing angles
$\theta_{12}$ and $\theta_{13}$ from individual   
observables.  

1). At present only the gallium experiments can be considered as the 
low energy experiments. For the Ge-production rate, $Q_{Ge}$, we have 
\be
Q_{Ge} = \left[ P_{ee}^{pp} Q_{Ge}^{pp} + P_{ee}^{pep} Q_{Ge}^{pep} + 
P_{ee}^{Be} Q_{Ge}^{Be} + P_{ee}^{CNO} Q_{Ge}^{CNO} +
P_{ee}^{B} f_B Q_{Ge}^{B} \right], 
\label{Gerate}
\ee
where $Q_{Ge}^{i}$ is the contribution to the rate from 
the
$i$th component of the solar neutrino flux  according to the SSM,
$P_{ee}^{i}$ is the corresponding effective (averaged over the energy range)
survival probability. 
The rate has the contributions both from the low 
and  high energy parts of the spectrum, though 
the latter (from boron neutrinos) is much smaller. We can subtract the 
boron neutrino effect, $Q_{Ge}^{B,exp}$,  using experimental results from 
SNO: $Q_{Ge}^l \equiv Q_{Ge} - Q_{Ge}^{B,exp}$. 
Taking the $^8{B}$ flux from SNO as an  input, we find that  
the $^8{B}$ neutrino contribution equals 
$Q_{Ge}^{B,exp} = 4.2 ^{+1.4}_{-0.7}$ SNU and then  
$Q_{Ge}^l = 63.9 ^{+5.1}_{-5.2}$ SNU for the rest of the neutrinos
\cite{ga}.

The conversion effect on $Q_{Ge}^l$ is described by the 
probability in low energy limit:
\be
Q_{Ge}^l = (Q_{Ge}^{SSM} - Q_{Ge}^{B})\left[\cos^4 \theta_{13}(1  - 0.5 
\sin^2 2\theta_{12})
- \cos^6 \theta_{13} \cos 2\theta_{12} \sin^2 2\theta_{12} 
\eta^{-1}_{Ge}\right],  
\label{germ}
\ee
where 
\be
\eta^{-1}_{Ge} \equiv \frac{1}{Q_{Ge}^{SSM} - 
Q_{Ge}^B}
\left(\eta^{-1}_{pp}Q_{Ge}^{pp} + 
\eta^{-1}_{pep}Q_{Ge}^{pep} +
\eta^{-1}_{Be}Q_{Ge}^{Be} +
\eta^{-1}_{CNO}Q_{Ge}^{CNO} \right).  
\ee

%%%%%%%%%%%%%%%%%ffff3%%%%%%%%%%%%%%%%%%%%%%%%%%%%%%%%%%%%%%%%%%%%%%%%%%%
\begin{figure}
\includegraphics[width=12.0cm, height=14cm]{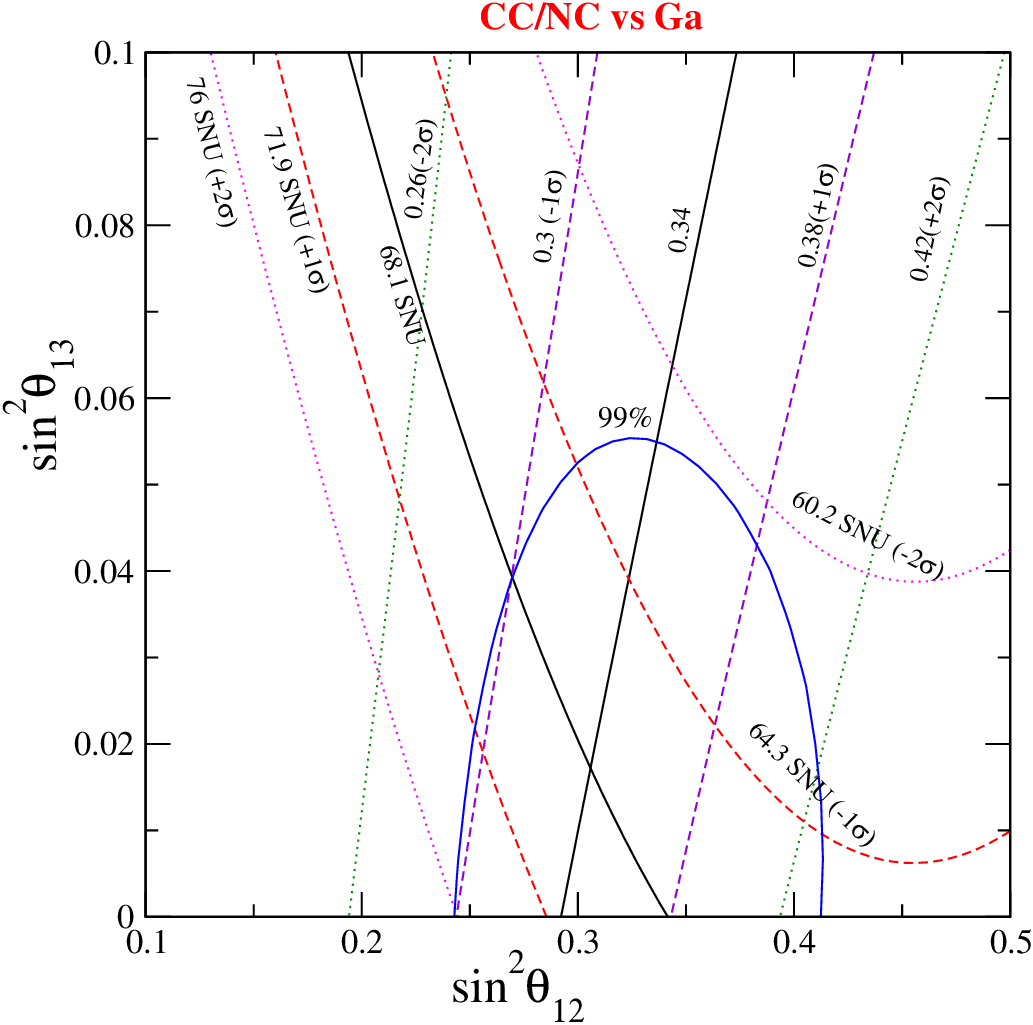}
\caption{The 
iso-contours  of the CC/NC ratio at SNO and Ge-production rate  
in the $\sin^2\theta_{12} -\sin^2\theta_{13}$ plane for 
$\dm = 8 \cdot 10^{-5}$ eV$^2$.
Shown are the iso-contours for the central  values of observables 
and  the $\pm 1\sigma$ and $\pm 2\sigma$ deviations. 
We take $f_B = 1$. 
The solid (blue) contour bounds the 99\% C.L. allowed area from the  
global fit of all data.
}
\label{ccbyncga}
\end{figure}
%%%%%%%%%%%%%%%%%%%%%%%%%%%%%%%%%%%%%%%%%%%%%%%%%%%

\begin{figure}
\includegraphics[width=12.0cm, height=14cm]{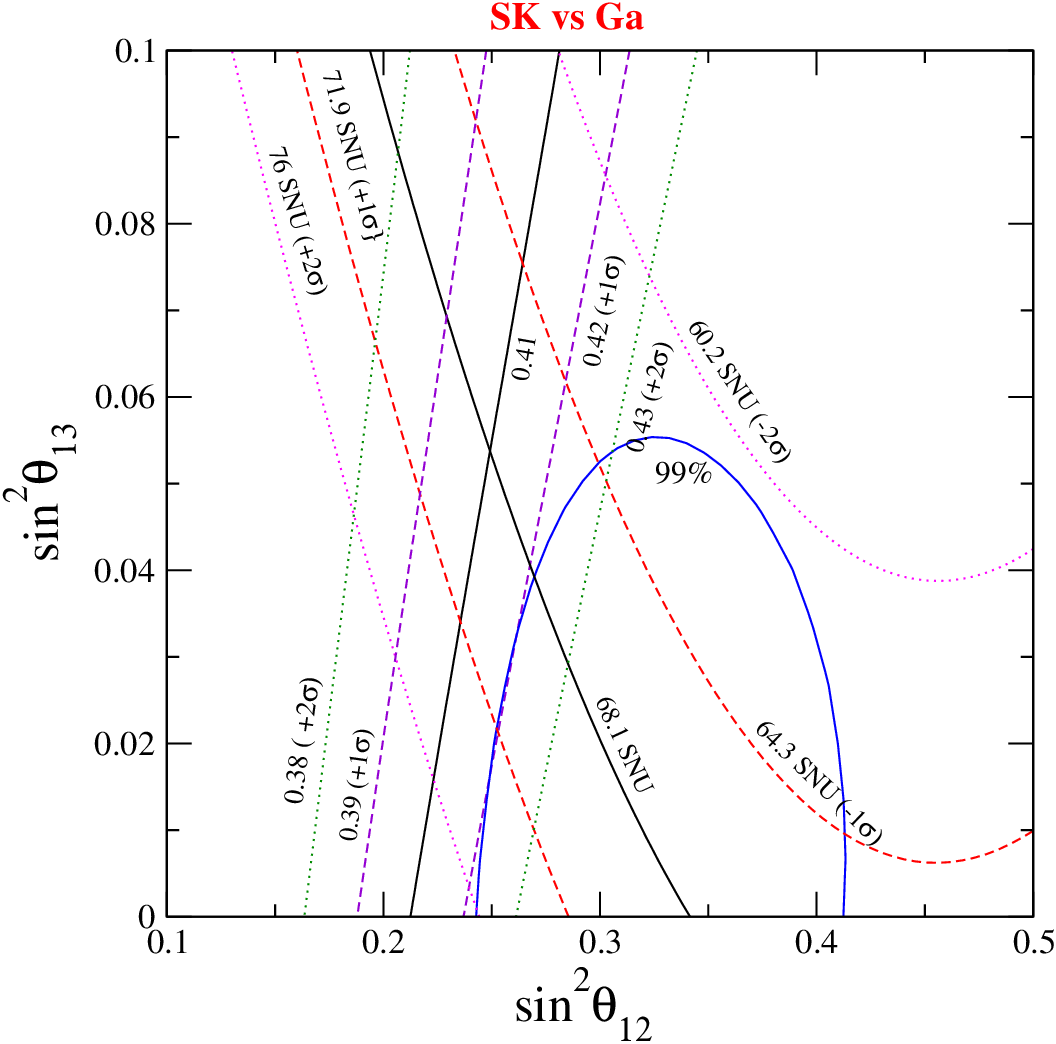}
\caption{The
iso-contours of  $R_{SK}$ and Ge-production rate   
in the $\sin^2\theta_{12} -\sin^2\theta_{13}$ plane for 
$\dm = 8 \cdot 10^{-5}$ eV$^2$.
Shown are the contours for the central values and $\pm 1 \sigma$,  
$\pm 2\sigma$ deviations from these central values. 
We take $f_B = 1$. 
The solid (blue) contour bounds the 99\% C.L. allowed area from the global
fit of all data.
}
\label{skvsga}
\end{figure}

\begin{figure}
\includegraphics[width=12.0cm, height=14cm]{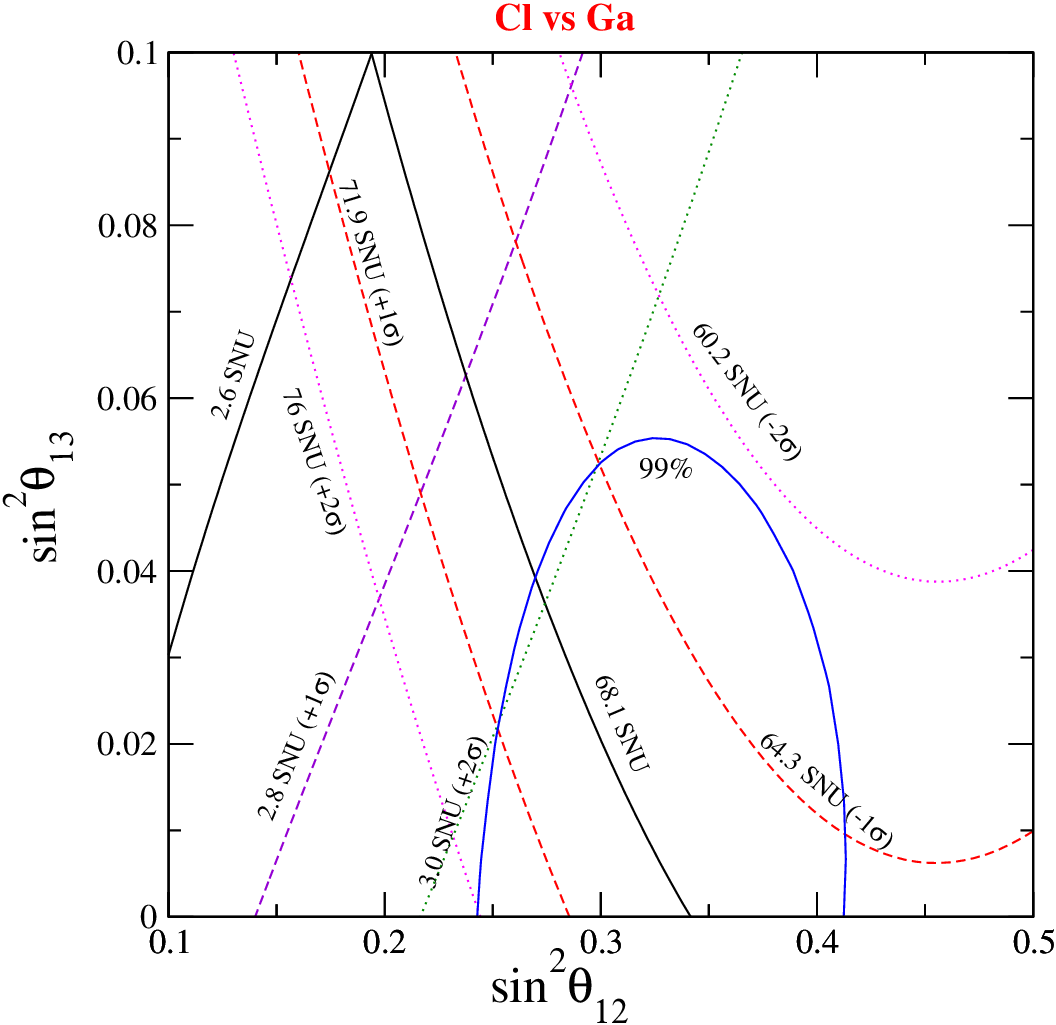}
\caption{The iso-contours of the Ar-   
and  Ge- production rates  
in the $\sin^2\theta_{12} -\sin^2\theta_{13}$ plane for 
$\dm = 8 \cdot 10^{-5}$ eV$^2$.
Shown are the contours for 
the central values as well as $\pm 1\sigma$ and $\pm 2\sigma$
deviations.  The contours have been calculated for $f_B = 1$. 
%The solid line restricts the 99\% C.L. allowed area from the  
%global fit of all data.
}
\label{isoclga}
\end{figure}

At the same time we find that dependence of the total rate 
$Q_{Ge}$ on $f_B$ and the high energy conversion effect 
is very weak: $10\%$ variations of $f_B$ would produce change of 
$Q_{Ge}$ about 0.4 SNU. 
Therefore  results of numerical calculations are given 
in figs.~\ref{ccbyncga}, \ref{skvsga},  \ref{isoclga},  
for $Q_{Ge}$ and $f_B = 1$.  
We show the contours of constant total rate 
$Q_{Ge}$  which correspond to the 
present central  value as well as to  $1\sigma$ and $2\sigma$ deviations 
from it. 
The contours are well described by the analytical expression given 
in (\ref{con-l}) with  $\eta^{-1}_{Y} =  \eta^{-1}_{Ge}$ and $C_Y \equiv 
C_{Ge}$. 

For zero 1-3 mixing the central value  $Q_{Ge} = 68.1$ SNU 
corresponds to $\sin^2\theta_{12} = 0.34$,  
and the best fit value {\bf{ $\sin^2\theta_{12} =  0.33$}} 
is accepted at $0.5\sigma$ level. 
For a given $Q_{Ge}$,  $\sin^2\theta_{13}$ increases 
with decrease of $\sin^2\theta_{12}$.  
Taking the best fit value  $\sin^2\theta_{12} = 0.33$ (see 
eq. (\ref{globs}))  
and $Q_{Ge} = 68.1$ SNU we find $\sin^2\theta_{13} = 0.004$. \\

2). The CC/NC ratio of events at SNO is given  
by the integrated (with cross-section) survival probability at high energies: 
CC/NC $= \langle P_{ee}^{(h)}\rangle$. 
The iso-contours are described approximately  by Eq. (\ref{con-h}) 
with $C_X \equiv C_{SNO} =$ CC/NC. 
The results of exact numerical calculations 
are shown in fig.~\ref{ccbyncga}. The iso-contours are constructed for the 
present central value of CC/NC as well as for  
its $1\sigma$ and $2\sigma$ deviations. 
At $\sin^2\theta_{12}=0.33$ the central value of CC/NC gives 
$\sin^2\theta_{13} = 0.046$.

As seen from fig.~\ref{ccbyncga}, 
the combination of SNO CC/NC and Ga-results prefers the non-zero 
but statistically insignificant 1-3 
mixing.  The iso-contours of the central observed  values 
of CC/NC and $Q_{Ge}$ cross at 
\be 
\sin^2\theta_{13} = 0.017 \pm 0.026  
\label{13dev}
\ee 
which is about $0.7\sigma$ deviation from zero. Here $1\sigma$ is 
evaluated from the fig.~\ref{ccbyncga}  
using CC/NC and $Q_{Ge}$ measurements.

The CC/NC ratio is free of the $^{8}$B neutrino normalization  
factor $f_B$. Iso-contours for the 
Ge-production rate are constructed for $f_B$=1,  {\it i.e.},  
for the SSM value of  the $^{8}$B flux. 
The dependence  of  $Q_{Ar}$  on $f_B$ essentially disappears if one uses 
for the boron neutrino flux the value measured by NC in SNO or 
by  SK. In any case the influence of uncertainty related to $f_B$  
on  the intersection of iso-contours is weak. 
Larger uncertainty ($\sim 1$ SNU) follows from the  Be-neutrino flux 
contribution. That leads to $\Delta \sin^2\theta_{13} \sim 0.01$. 
Future Borexino measurements can reduce this uncertainty. \\

3). The SuperKamiokande rate of the  $\nu e-$events. 
The ratio of the observed to SSM predicted event rates is given by 
\be
R_{SK} = f_B [\langle  P_{ee} \rangle (1 -  r) + r],
\label{rsk}
\ee
where  $r \equiv \sigma_{\nu_{\mu}}/\sigma_{\nu_e}$ is the ratio of 
the $\nu_e - e$ and $\nu_{\mu} - e$ cross-sections. 
The lines of constant $R_{SK}$ coincide with the lines 
$\langle  P_{ee} \rangle = C_{SK} =  const$, where according to 
(\ref{rsk})
\be
C_{SK} = \frac{R_{SK}/f_B - r}{1 - r}. 
\label{lines}
\ee
The lines are given by the  Eq. (\ref{con-h}) 
with $C_X =  C_{SK}$ and the 
averaging (integration) $\langle ... \rangle_{X}$ should  be 
done according to the SK experiment characteristics. 

%%%%%ffff4%%%%%%%%%%%%%%%%%%%%%%%%%%%%%%%%%%%%%%%%%%%%%%%%
%%%%%%%%%%%%%%%%%%%%%%%%%%%%%%%%%%%%%%%%%%%%%%%

The iso-contours of the SK rate in the $\sin^2\theta_{12} 
-\sin^2\theta_{13}$ plane are presented in fig.~\ref{skvsga} for $f_B = 
1$.   They  correspond to the SK central value 
$R_{SK} =  0.41$ as well as the  1$\sigma$ and 2$\sigma$ deviations.  
According to  fig.~\ref{skvsga}, for $\sin^2\theta_{13} =0$  
the  ratio  $R_{SK} =  0.41$ is achieved at   
$\sin^2\theta_{12} \approx 0.21$ which is about $2\sigma$ 
below the best global value.  
In turn,  the best global value  $\sin^2\theta_{12} = 0.33$ 
and $R_{SK} =  0.41$ lead to $\sin^2\theta_{13}= 0.17$.  

With decrease of $f_B$ the corresponding iso-contours shift to 
larger $\theta_{12}$ and smaller $\theta_{13}$. For arbitrary $f_B$ 
one can still use 
the grid of contours calculated for $f_B = 1$, but with changed  
values $R_{SK}^{(1)}$. 
Indeed, according to (\ref{lines}), for given values $R_{SK}$ 
and $f_B$ one should take the contour $R_{SK}^{(1)} = R_{SK}/f_B$.

The combination of SK- and  Ga- data  
(crossing point of the contours which correspond 
to the central  experimental values) selects  $\sin^2\theta_{13} =  0.053$  
as is seen from fig.~\ref{skvsga}. 
For $f_B = 0.84$ the best fit SK value, $R_{SK} = 0.41$, 
corresponds  to the iso-contour with $R_{SK}^{(1)} = 
0.49$. The intersection of this contour 
with the iso-contour $Q_{Ge} = 68.1$ SNU occurs at 
much smaller 1-3 mixing: $\sin^2\theta_{13} =  
- 0.004$.\\ 
%%and $\sin^2\theta_{12} =  0.33$.\\ 

4). The Argon production rate in the Homestake experiment  is 
determined by 
\be
Q_{Ar} = \left[ 
P_{ee}^{Be} Q_{Ar}^{Be} + P_{ee}^{pep} Q_{Ar}^{pep} + 
P_{ee}^{CNO} Q_{Ar}^{CNO} + P_{ee}^{B} f_B Q_{Ar}^{B} \right]. 
\label{Arrate}
\ee 
According to  the SSM $Q_{Ar}$
is dominated by the high energy (boron) neutrinos. 
Therefore the contours of constant rate (fig. \ref{isoclga})
have typical slope  of the high energy observables.  
We show these contours for $f_B = 1$.

%%%%%%%%%%%%%%%%%%%%%%%%%%ffff5%%%%%%%%%%%%%%%%%%%%%%%%%%%%%%%%%
%%%%%%%%%%%%%%%%%%%%%%%%%%%%%%%%%%%%%%%%%%%%%%%%%%%%%%%%%%

For $\sin^2\theta_{13} =0$
and the best global value of $\sin^2\theta_{12}$ the rate  
$Q_{Ar}$  is  about 3.15 
SNU which is $2.5\sigma$ above the experimental result.  
The combination of  Cl-  and  Ga-results prefers  large  
1-3 mixing: $\sin^2\theta_{13} = 0.1$ which deviates from 
zero by $\sim 2.5\sigma$.  

For the arbitrary $f_B$ again one needs to rescale 
$Q_{Ar}^{B} \rightarrow Q_{Ar}^{B}/f_B$. Thus, for $f_B = 0.84$ 
and $Q_{Ar} = 2.55$ SNU the iso-contour with  
$Q_{Ar} \approx  3.04 $ SNU should be used. It intersects with 
iso-contour $Q_{Ge} = 68.1$ SNU at $\sin^2\theta_{13} = 0.032$.\\

Summarizing, 
for $\Delta m^2_{21} = 8 \cdot 10^{-5}$ eV$^2$, the     
combination of SNO and Ga data selects $\sin^2\theta_{13} = 0.017$, 
which is practically independent of $f_B$.   
For $f_B = 0.84$ (which follows from the global fit)  
the combination of SK and Ga results leads to $\sin^2\theta_{13} = - 0.004$ 
and the combination of Cl and Ga data selects 
$\sin^2\theta_{13} \sim 0.032$, however statistical 
significance of the latter is low. 
The average from these three combinations is about 
$\sin^2\theta_{13} \sim 0.01$. It agrees 
well with  results of fit of all the solar data 
for the same   $\Delta m^2_{21}$. For larger $f_B$ the larger values of  
$\sin^2\theta_{13}$ would be preferable. 

Notice that available  spectral information does not give strong  
restriction on $\sin^2\theta_{13}$.  
In fact, flatness of the 
measured spectrum of events at SNO and SK may even favor non-zero  
$\sin^2\theta_{13}$. 
Indeed, at high energies an increase of $\sin^2\theta_{13}$ implies 
increase of $\sin^2\theta_{12}$ and therefore flattening of the spectrum. \\

The results on 1-3 mixing described above  are in agreement with 
the best fit value from the global analysis of all solar 
neutrino data, $\sin^2\theta_{13} = 0.01$,  
which is realized, however, for smaller $\Delta m^2_{21}$ (\ref{globs}). 
Inclusion of the CHOOZ result  diminishes  
$\sin^2\theta_{13}$ to practically zero.\\ 

Let us consider  dependence of the iso-contours on  
$\Delta m^2_{21}$. Using (\ref{con-h}) we find for the high energy part: 
\be
\delta (\sin^2 \theta_{13}) \approx
\frac{1}{2} \left[\sin^2 \theta_{13} + \frac{C_X - \sin^2\theta_{12}   
-3\langle f_{reg}\rangle_{X}(0.5 - \sin^2\theta_{13})}{
\sin^2\theta_{12} + 1.5 \langle f_{reg} \rangle_{X}} \right] \delta_{\Delta},
\label{var-h}
\ee
where we have expressed $\eta$ in terms of $\sin^2 \theta_{13}$. 

In the low energy region according to (\ref{con-l}) 
\be
\delta(\sin^2 \theta_{13})  \approx
\frac{\eta_{Y}^{-1}}{4} 
\frac{\cos 2\theta_{12} \sin^2 2\theta_{12}}{1 
- 0.5 \sin^2 2\theta_{12}}~ \delta_{\Delta}, 
\label{con-l2}
\ee
and the  pre-factor  is small: $\delta(\sin^2 \theta_{13}) \approx 0.011 
\delta_{\Delta}$.  

\begin{figure}
\includegraphics[width=12.0cm, height=14cm]{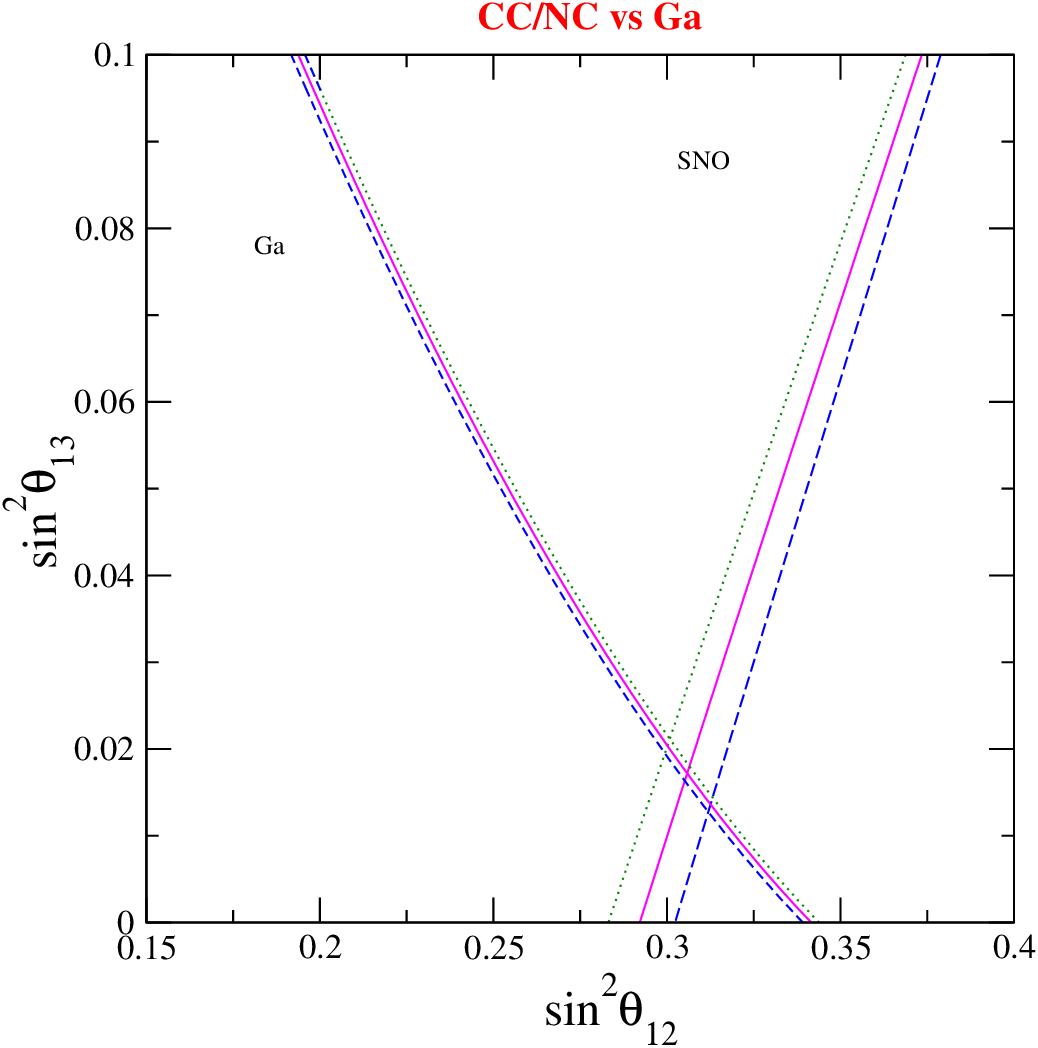}
\caption{The
iso-contours of  CC/NC $= 0.34$ 
at SNO and $Q_{Ge} = 68.1$ SNU in Ga experiments
in the $\sin^2\theta_{12} -\sin^2\theta_{13}$ plane for different values 
of $\dm$: 
$\dm = 9 \cdot 10^{-5}$ eV$^2$ - the dotted lines;  
$\dm = 8 \cdot 10^{-5}$ eV$^2$ - the solid lines; 
$\dm = 7 \cdot 10^{-5}$ eV$^2$ - the dashed lines. 
}
\label{ccbyncvsga_delmsq}
\end{figure}

In fig.~\ref{ccbyncvsga_delmsq} we show results of numerical 
computations  
of the iso-contours for Ga-rate  and CC/NC ratio for three different 
values of $\Delta m^2_{21}$. In agreement with our qualitative
consideration a  shift of the low energy contours is much weaker. 
Variations of $\Delta m^2_{21}$ from $7$ to $9 \cdot 10^{-5}$ 
eV$^2$   (which is about 25\% change) gives 
$\delta(\sin^2 \theta_{13}) = 0.008$ and 
$\delta(\sin^2 \theta_{12}) = 0.012$. 
Reduction of the uncertainty in $\Delta m_{21}^2$ down to 14\% will result 
in $\delta(\sin^2 \theta_{13}) = 0.004$ and 
$\delta(\sin^2 \theta_{12}) = 0.007$. 
This is much smaller than the expected sensitivity of future solar 
neutrino studies. The contours have been obtained for 
$f_B = 1$. We have found that $10\%$ variations of $f_B$ 
produce  shift of the Ge-lines by $\delta(\sin^2 \theta_{12}) = 0.004$, 
and the SNO contours do not change at all.\\

Let us consider the effect of matter corrections to the 
1-3 mixing. As we have established in sec. 2, the effect is reduced to 
renormalization of $\sin^2 \theta_{13}$ according to Eq. (\ref{sin2}). 
True value of $\sin^2 \theta_{13}$ is smaller than the one extracted from the data 
without taking into  account matter correction. The 
renormalization  is negligible for the low energy observables and it 
is of the order 5\% at high energies. So, the iso-contours of 
high energy observables should be shifted down by factor 
$\sim 0.95$. This, in turn leads to decrease of 
$\sin^2 \theta_{13}$ in the intersection points:  {\it e.g.}, instead of 
$\sin^2 \theta_{13} = 0.017$ one finds 0.016.

%figure 7 : ccbync + ga different delta msq
%%%%%%%%%%%%%%%%%%%%%%%%%%ffff6%%%%%%%%%%%%%%%%%%%%%%%%%%%%%%%%%%%%%%%%%%
%%%%%%%%%%%%%%%%%%%%%%%%%%%%%%%%%%%%%%%%%%%%%%%%%%%%%%%%%%%

The  figures presented in this section 
allow us to estimate relevance of the non-zero 1-3 mixing for the 
solar neutrino.

For a given  CC/NC ratio 
an increase of $\sin^2\theta_{13}$ from 0 to 0.05 
($\sim 3\sigma$ upper bound) corresponds to decrease 
the  Ge-production  rate by $2\sigma$ 
or $\Delta Q_{Ge} \sim - 8$ SNU.  Inversely, for a given $Q_{Ge}$, 
the same increase of $\sin^2\theta_{13}$  
leads to $1.8\sigma$ decrease of the ratio CC/NC: 
$\Delta$ CC/NC $\sim -0.05$. 
The same change of $\sin^2\theta_{13}$ 
diminishes the SK rate  by $4\sigma$: $\Delta R_{SK} = -0.04$. 
So, the $1 \sigma$ variations of $\theta_{13}$ produce about 
$1 \sigma$ or smaller variations of observables. 
As we have discussed in the beginning 
of previous section, the impact of $\theta_{13}$ on the global fit of the 
solar neutrino data is much weaker: $3\sigma$ increase of 
$\sin^2\theta_{13}$  leads to only $0.7\sigma$ increase of 
$\sin^2\theta_{12}$.

%%%%%%%%%%%%%%%%%%%%%%%%%%%%%%%%%%%%%%%%%%%%%%%%%%%%%%%%%%%%%%%%%%%
\section{Sensitivity of the  future solar 
neutrino experiments to 1-3 mixing} 
%%%%%%%%%%%%%%%%%%%%%%%%%%%%%%%%%%%%%%%%%%%%%%%%%%%%%%%%%%%%%%%%%%%%

As we have shown, some sensitivity of the solar neutrino data 
to the 1-3 mixing appears essentially due 
to the combination of  the SNO- and SK- results from the one hand side and 
the Gallium results from the other side. 
Future high precision measurements will have better sensitivity. 
To evaluate this sensitivity and to 
study possible implications of measurements we have constructed 
fine grids of the iso-contours of various observables. 
We will use the expected $1\sigma$ errors, $\sigma_{13}$, in 
determination of $\sin^2 \theta_{13}$ as a measure of the sensitivity. 
$2\sigma$ and $3\sigma$  errors can also be estimated from the 
grids of iso-contours we present. In the first approximation 
$n\sigma$ error is given by $n \sigma_{13}$. \\

%%
%%%%%%%%%%%%%%%%%%%%%%%%%%%%%%%%%%%%%%%%%%%%%%%%%%%%%%%%%%%%%%%%%
%figure 8: ccbync+ga_reduced
%%%%%%%%%%%%%%%%%%%%%%%%%%ffff7%%%%%%%%%%%%%%%%%%%%%%%%%%%%%%%%%%%%%%%%%%%
\begin{figure}
\includegraphics[width=12.0cm, height=14cm]{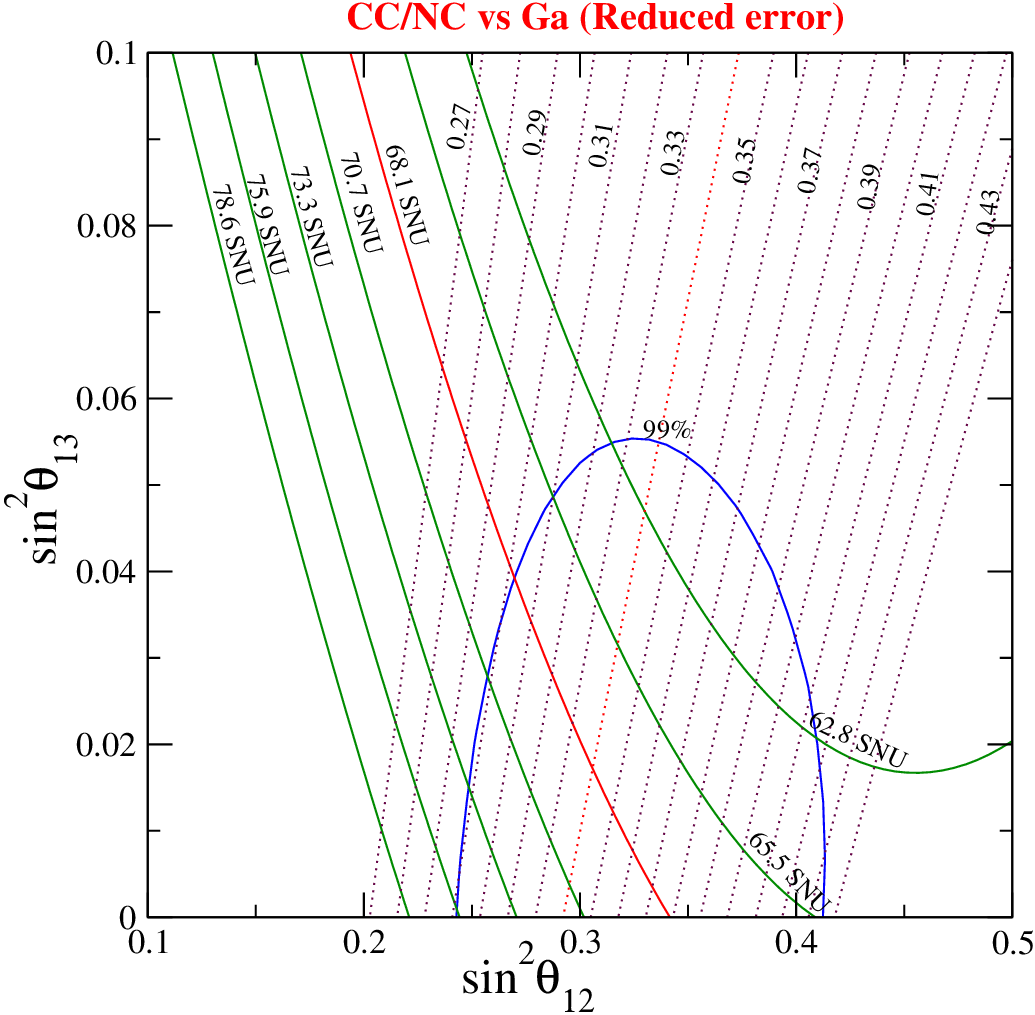}
\caption{The iso-contours of the  CC/NC ratio 
at SNO (dotted lines) and  Ge-production rate $Q_{Ge}$ (solid line)
with reduced steps  0.01 and 2.6 SNU respectively, 
in the $\sin^2\theta_{12} -\sin^2\theta_{13}$ plane.
$\dm$ is held fixed at $8 \cdot 10^{-5}$ eV$^2$.
%The solid (blue)  contour gives the 99\% C.L. allowed area from the 
%present global data.
}
\label{ccbyncga_reduced}
\end{figure}
%%%%%%%%%%%%%%%%%%%%%%%%%%%%%%%%%%%%%%%%%%%%%%%%%%%%%%%%%%%%%%%%%%%
%%
1). In fig.~\ref{ccbyncga_reduced} we show the 
fine grid  of iso-contours   for the CC/NC-ratio 
and the Ge- production rate. 
The SNO (phase III) will measure the NC- event rate with  
6\% accuracy \cite{sno3}. 
If we assume a reduced error of 5\% for the CC-data then this makes the 
CC/NC error as 8\% which can be  compared to the present error 
of 11\%.  (This corresponds to  reduction of the absolute error in the 
CC/NC ratio  down to 0.024  from the current value of 0.035.) 

According to fig.~\ref{ccbyncga_reduced},  
the current  Ge-production rate accuracy is  
$\Delta Q_{Ge} \approx 4$  SNU.  
Using this accuracy and the expected SNO-III error,  
we find from fig.~\ref{ccbyncga_reduced} the 
sensitivity ($1\sigma$ error) to $\sin^2\theta_{13}$ in the region 
$\sin^2\theta_{13} \sim 0 - 0.03$: 
\be
\sigma_{13}({\rm CC/NC} + {\rm Ga}) = 0.023.
\ee
It is  only slightly better than the present one (\ref{13dev}).

If we take 5\%  accuracy for future possible measurements of 
CC/NC ratio and the Ge-production rate error 2.6 SNU 
(the later is slightly larger than the present systematic error 
\cite{ga}, and it could correspond to future hypothetical large mass  
Gallium experiment for which statistical error is substantially 
smaller than the systematic one; this can be considered as 
the ultimate sensitivity of the Gallium experiments), 
then according to fig.~\ref{ccbyncga_reduced}  
\be
\sigma_{13}({\rm CC/NC} + {\rm Ga}) = 0.016.
\ee
In this case the value $\sin^2 \theta_{13} = 0.017$ would have 
only $1 \sigma$ significance, and in the case of very small 
$\theta_{13}$  the upper bound will 
be $\sin^2 \theta_{13} < 0.024 ~(0.048)$
at $90\%~~ (3\sigma)$. \\

2). The Borexino and KamLAND plan to measure   
signal from the $^7 Be$ neutrinos.
The iso-contours can be well described by the low energy formula  
(\ref{con-l}). Taking into account contribution from the 
non-electron (converted) neutrinos we find  
the  iso-contours, 
$R_{Borexino}$, which coincide  with iso-contours 
$\langle  P_{ee} \rangle = 
C_{Be}$,  where $C_{Be} = (R_{Borexino} - r)(1 - r)^{-1}$. 

%%%%%%%%%%%%%%%ffff8 %%%%%%%%%%%%%%%%%%%%%%%%%%%%%%%%%%%%%%%%%%%%%%%%%%
\begin{figure}
\includegraphics[width=12.0cm, height=14cm,clip=]{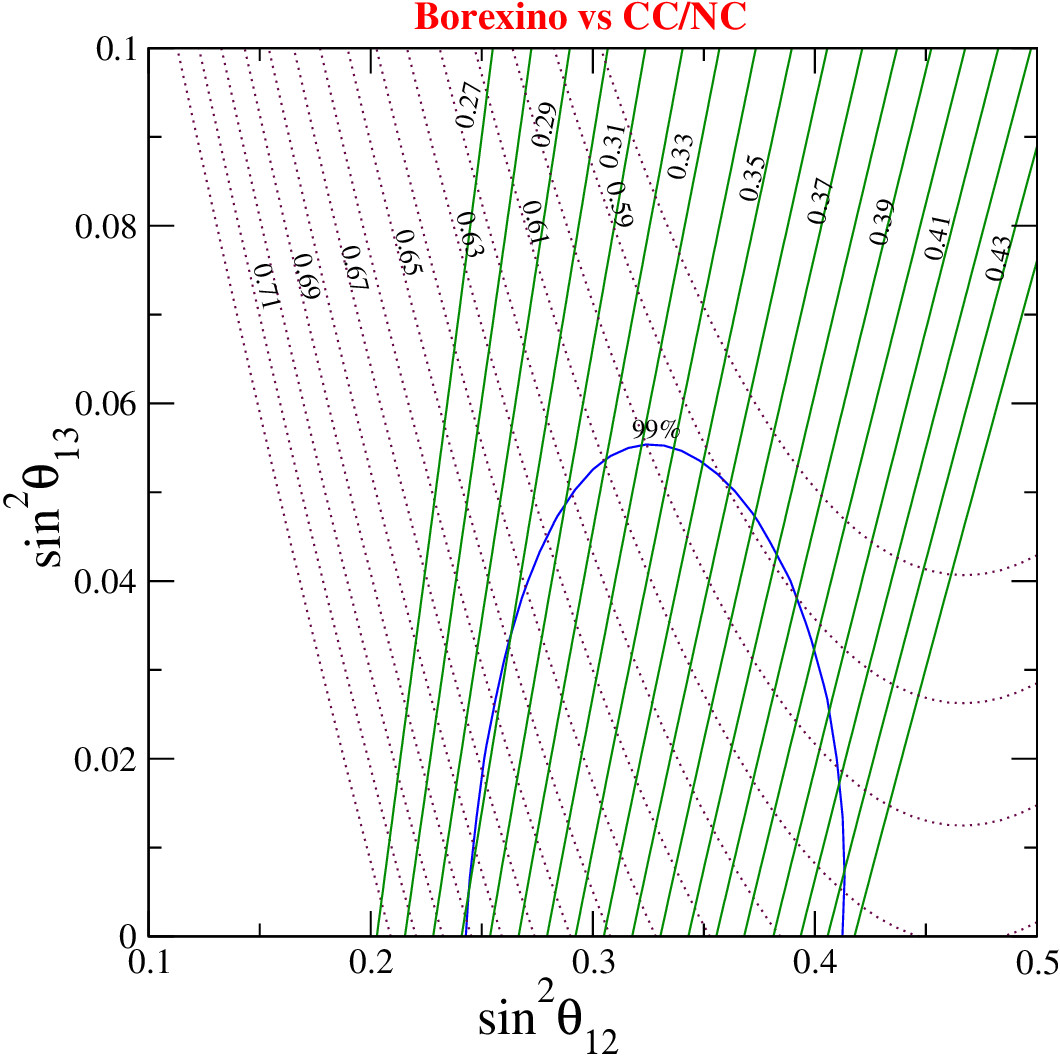}
\caption{The iso-contours of Borexino rate suppression with respect to 
the SSM prediction in the $\sin^2\theta_{12} -\sin^2\theta_{13}$ plane for 
$\dm = 8 \cdot 10^{-5}$ eV$^2$. 
%The solid (blue)  contour gives the 99\% C.L. allowed area from the 
%global fit of all data.
}
\label{isoborex}
\end{figure}
%%%%%%%%%%%%%%%%%%%%%%%%%%%%%%%%%%%%%%%%%%%%%%%%%%%%

The results of calculations  are shown in  fig. \ref{isoborex}.   
The Borexino contours have similar form to those for 
Ga-rate. 

Borexino is expected to accumulate 44000 events in 4 years of time 
($\sim$ 30 events/day) for LMA MSW solution. This corresponds to a
statistical error of only 0.5\% \cite{borexnu2004}. But  
the accuracy of measurement will be  dominated   by the 
background and systematics.  
Furthermore, the SSM prediction for the $^7 Be$ flux has 10\% uncertainty  
which will  add to the total error  unless one keeps 
the normalization of $^7 Be$ flux as a free parameter 
to be determined from the Borexino experiment itself with a higher 
accuracy. 
Using fig.~\ref{isoborex}  we estimate that  
a 5\% accuracy of measurements of the rate at Borexino  
will contribute to improvement of  
sensitivity of the solar studies  to $\sin^2\theta_{13}$.\\ 

3). New low energy solar neutrino experiments  
will be able to measure the pp-neutrino flux  with 
high accuracy~\cite{ppex}. 
Two kinds of experiments are being discussed~\cite{lownu}:  
one -- using the charged
current reactions (LENS, MOON, SIREN) and the other -- using the  
neutrino-electron scattering process (XMASS, CLEAN, HERON, MUNU, GENIUS). 

For the pp-neutrinos we can use 
the iso-contour equation for the low energies   
as in (\ref{con-l}) with 
substitution $C_Y = C_{pp} = \langle  P_{ee} \rangle_{pp}$  
and $\eta^{-1}_{pp} = \langle E V/ 
\Delta m^2_{21} \rangle_{pp}$.  Practically,    $\eta^{-1}_{pp}$ is 
negligible. 
%%
%%%%%%%%%%%%%%%%%%%%ffff9%%%%%%%%%%%%%%%%%%%%%%%%%%%%%%%%%%%%%%%%%%%%%%%%%%
%Figure 2: isopp vs CC/NC
%%%%%%%%%%%%%%%%%%%%%%%%%%%%%%%%%%
\begin{figure}
\includegraphics[width=12.0cm, height=14cm]{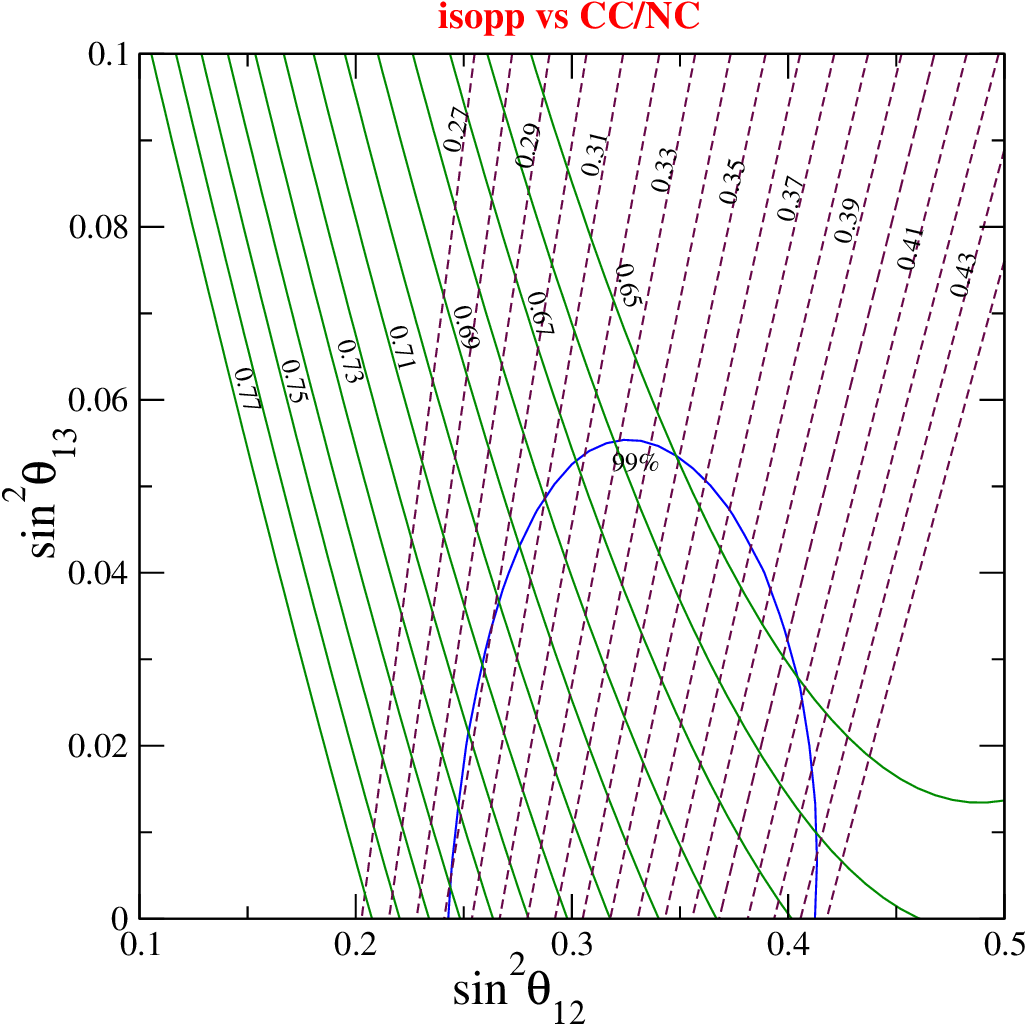}
\caption{The
iso-contours of  the CC/NC ratio (dashed lines) and   
$\nu e$ scattering rate of pp-neutrinos 
in the $\sin^2\theta_{12} -\sin^2\theta_{13}$ plane for 
$\dm = 8 \cdot 10^{-5}$ eV$^2$.
%The solid (blue)
%contour gives the 99\% C.L. allowed area from the  global
%fit of all data.
}
\label{ccbyncpp}
\end{figure}
%%%%%%%%%%%%%%%%%%%%%%%%%%%%%%%%%%%%%%%%%%%%%%%%%%%%%%%%%%%%%%%%%%%%%%%%%%%%%%%%%

In fig.~\ref{ccbyncpp} we show the contours of constant  
rate of the $\nu-e$ scattering events due to the 
pp-neutrinos. The neutral current contribution is also included. 
According  to fig.~\ref{ccbyncpp},  future measurements
of the CC/NC ratio with 5\% accuracy  and   the pp-rate   with 2\% accuracy 
will have a sensitivity  ($1\sigma$ error)
\be
\sigma_{13}({\rm CC/NC} + pp) = 0.015.
\label{senscc}
\ee
Consequently, the value $\sin^2 \theta_{13} = 0.034$ will be established  
with $2.3\sigma$ significance,  and in the case of very small 
$\theta_{13}$
the upper bounds  $\sin^2 \theta_{13} < 0.022 ~(0.045)$
at $90\%~~ (3\sigma)$ can be achieved.\\

%%%%%%%%%%%%%%%%%%%%%%%%%%%%%5%%%%%%%%%%%%%%%%%%%%%%%%%%%%%%%%%%%

\begin{figure}
\includegraphics[width=12.0cm, height=14cm]{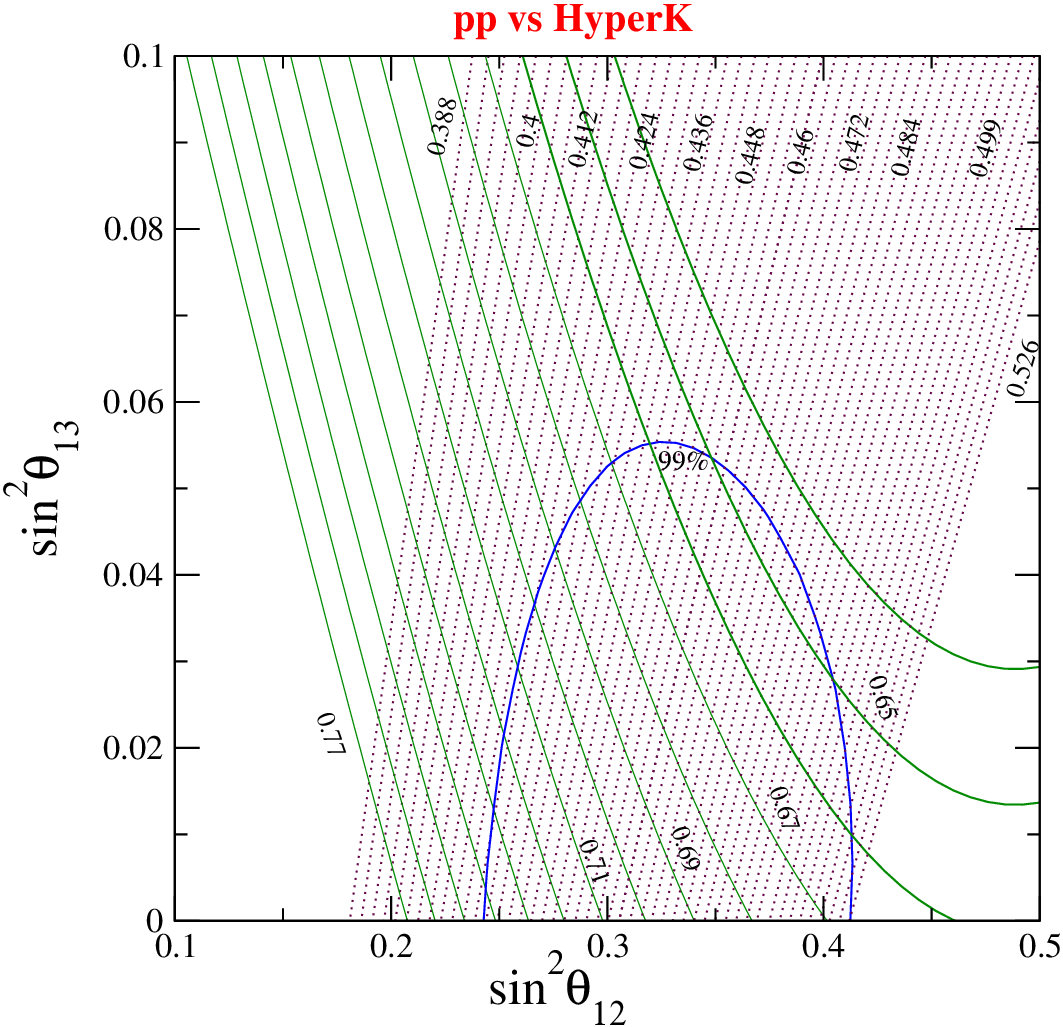}
\caption{The iso-contours  of  the neutrino 
electron scattering rate for the  pp-neutrinos
(solid line) and high energy $^8B$ neutrinos (dotted line)
in the $\sin^2\theta_{12} -\sin^2\theta_{13}$ plane.
The iso-rates for the high energy neutrinos are drawn
with an step 
0.003 corresponding to the expected statistical error of a Megaton water  
detector like HyperKamiokande.
$\dm$ is held fixed at $8 \cdot 10^{-5}$ eV$^2$ and $f_B = 1$. 
%The solid (blue)  contour gives the 99\% C.L. allowed area from the 
%present global fit of data.
}
\label{ppvshyperk}
\end{figure}

4). Future Megaton scale water Cherenkov detectors
HyperKamiokande \cite{hyperk} and  UNO \cite{UNO}   
will be able to measure the ratio $R_{\nu e}$ with accuracy 
$\Delta R_{\nu e} \sim 0.003$. We construct the fine grid of 
the $R_{\nu e}$ iso-contours  for $f_B = 1$ (fig.~\ref{ppvshyperk}).  
According to fig.~\ref{ppvshyperk} these measurements of $R_{\nu e}$
and  the pp-rate   (2\% accuracy) 
would have a sensitivity  ($1\sigma$ error)
\be
\sigma_{13}({\rm CC/NC} + pp) = 0.011.
\label{sensnue}
\ee
However,  the problem here is the poor knowledge of  $f_B$.  
It is difficult to expect that  accuracy of the theoretical predictions 
will be better than 10\%.  Comparable accuracy will be 
achieved in the direct measurements of the neutral currents.  
The global fit of the data which include the spectral information and 
the regeneration effect with free $f_B$ may give better accuracy. 

So, we conclude that the future solar neutrino studies 
may reach a sensitivity $\sin^2 \theta_{13} \sim 0.015 - 0.020$  
($1\sigma$), as compared with the present sensitivity 
0.05 - 0.06. It seems that even combined fit of several future 
precision measurements will not go down to 0.01 which may be   
considered as an ultimate sensitivity 
of solar neutrino measurements of  $\sin^2 \theta_{13}$. \\

%%%%%%%%%%%%%%%%%%%%%ffff10%%%%%%%%%%%%%%%%%%%%%%%%%%%%%%%%%%%%%%%%%%
%%%%%%%%%%%%%%%%%%%%%%%%%%%%%%%%%%%%%%%%%%%%%%%%%%%%%%%%%%%%%%%%%%%%%

On the basis of studies performed in secs. 5 and 6 we can 
identify signatures of the non-zero  $\sin^2\theta_{13}$. 
In general, they show up as a mismatch of the low and 
high energy measurements. The signatures include  

\begin{itemize}

\item 

Small value of the Ge production rate $Q_{Ge}$. For a given  ratio 
CC/NC one can find  the critical value of $Q_{Ge}^c({\rm CC/NC})$,   
so that inequality 
\be
Q_{Ge} < Q_{Ge}^c({\rm CC/NC})
\label{dif1}
\ee
will testify for the non-zero  value of $\sin^2\theta_{13}$. 
From the fig. \ref{ccbyncga_reduced} we find 
$Q_{Ge}^c(0.30) = 76$ SNU, $Q_{Ge}^c(0.32) = 74$ SNU, 
$Q_{Ge}^c(0.34) = 72$ SNU,  $Q_{Ge}^c(0.36) = 70$ SNU, {\it etc.}.
The larger the difference in (\ref{dif1}), the larger $\sin^2\theta_{13}$. 
In fact,  the present data satisfy  the inequality (\ref{dif1}), 
however the statistical significance is low. 
Apparently if CC/NC and $Q_{Ge}$ are measured with high  accuracy, 
the presence of non-zero $\sin^2\theta_{13}$ can be established.  

The inequality (\ref{dif1}) can be inverted,  so that 
small values of CC/NC for a given $Q_{Ge}$,   
\be
{\rm CC/NC} < {\rm CC/NC}^c(Q_{Ge}), 
\label{diff2}
\ee
testify for non-zero $\theta_{13}$.

\item 

Small Borexino rate:   
\be 
R_{Bor} < R_{Bor}^c({\rm CC/NC}). 
\ee
The critical Borexino rates for different values of CC/NC equal: 
$R_{Bor}^c(0.280) = 0.71$, $R_{Bor}^c(0.305) = 0.69$,
$R_{Bor}^c(0.325) = 0.67$, $R_{Bor}^c(0.340) = 0.66$,  $R_{Bor}^c(0.36) = 0.645$.

\item

Small $\nu e$ event rate produced by the pp-neutrinos:
\be
R_{pp} < R_{pp}^c({\rm CC/NC}). 
\label{minc}
\ee 
According to fig.~\ref{ppvshyperk} the critical values 
$R_{pp}^c({\rm CC/NC})$ 
equal 0.75 (0.294), 0.73(0.318), 0.71(0.343), 0.69(0.373).

\item 
Weak spectrum distortion of the boron neutrinos. The spectrum 
distortion  can be characterized, {\it e.g.}, by 
the ratio of suppressions at low and hight energies 
$r_R \equiv R_{\nue}(E < 7 ~{\rm MeV})/ R_{\nue}(E > 7~ {\rm MeV})$.  
There 
is certain critical value of the distortion,  $r_R^c$, 
which depends on  CC/NC, so that weaker distortion will testify 
for non-zero 1-3 mixing. Indeed, for fixed  CC/NC with increase of
$\sin^2 \theta_{13}$,  the 1-2 mixing, $\sin^2 \theta_{12}$,  should 
increase (\ref{highl3}) 
and therefore, the up turn of spectrum at low energies should be weaker.

\end{itemize}

%%%%%%%%%%%%%%%%%%%%%%%%%%%%%%%%%%%%%%%%%%%
\section{Implications of the solar neutrino measurements of $\theta_{13}$}
%%%%%%%%%%%%%%%%%%%%%%%%%%%%%%%%%%
%Figure 9

Let us first compare  sensitivities   of the future solar neutrino 
experiments and  future  reactor and accelerator experiments 
given in eqns. (\ref{th13lbl}, \ref{jparc}, \ref{dchooz}).  
With the expected accuracy of SNO-III and 
2\% accuracy of the pp-neutrino flux measurements  
the solar neutrino 
experiments can reach sensitivity comparable to the
one  of the forthcoming MINOS and CNGS experiments. 
This sensitivity is at the level of the  present bound from the global 
fit of all oscillation data and it is certainly lower  than  the 
sensitivity of future  measurements.

The solar neutrino sensitivity would be comparable to that of J-PARC and 
Double-CHOOZ if the 
pp-neutrino flux is measured with 1\% accuracy and CC/NC ratio - with 3\% 
accuracy which looks rather unrealistic now. 

Notice that the survival probability $P_{ee}$ does not depend on  
the CP phase $\delta$,  and therefore the problem of  degeneracy 
with $\delta$  \cite{barger} does not arise. So,  
the solar neutrinos studies can provide a clean channel for 
determination of $\theta_{13}$ like the reactor neutrinos.
Therefore, in spite of the lower sensitivity the solar neutrino data 
will contribute  somehow to resolution of degeneracy of these parameters 
in the future global fits.

It seems that the reactor and accelerator experiments 
will obtain results much before future high statistics solar neutrino 
experiments will start to operate. In this case  possible implications of 
the solar 
neutrino measurements  can include 

1). Precise determination of $\sin^2\theta_{12}$: 
resolution of the $\theta_{12} - \theta_{13}$ degeneracy.

2). Comparison of results on  $\theta_{13}$ (as well as on $\theta_{12}$)
from the solar neutrinos and reactor/accelerator experiments. 
The outcome could be further confirmation of the reactor/accelerator 
results by solar neutrinos, or establishing certain discrepancy  
which may imply new physics beyond the LMA solution of the solar neutrino 
problem. 

Let us consider these two points in some details. \\

%%%%%%%%%%%%%%%%%%%%%%%%%%%%%%%%%%%%%%%%%%%%%%%%%%%%
{\it 1). Resolving  degeneracy of angles.}  
%%%%%%%%%%%%%%%%%%%%%%%%%%%%%%%%%%%%%%%%%%%%%%%%%%%%
Due to the strong $\sin^2\theta_{12} - \sin^2\theta_{13}$  
degeneracy   
unknown value of  $\theta_{13}$ leads to uncertainty in 
the determination of $\sin^2\theta_{12}$ and {\it vice-versa}.  
Thus,  in the CC/NC measurements, inclusion of the 1-3 mixing produces an  
increase of $\sin^2\theta_{12}$ by $0.04 - 0.05$ (that is,  by 20\%) 
if $\sin^2\theta_{13}$ increases from 0 to 0.06
(see fig. \ref{ccbyncga_reduced}).  
In the   pp-neutrino flux measurements  $\sin^2\theta_{12}$ decreases by 
$0.06 - 0.08$,  when $\sin^2\theta_{13}$ increases from 0 to 0.06 (see 
fig.~\ref{ccbyncpp}). 
So,  precise determination of $\sin^2\theta_{12}$
($\sin^2\theta_{13}$)  from the solar neutrino data 
is not possible unless $\sin^2\theta_{13}$ ($\sin^2\theta_{12}$)
is measured or 
strongly restricted. 
As follows from our analysis a combination of high statistics measurement
at high and low energies can remove the degeneracy. \\

%%%%%%%%%%%%%%%%%%%%%%%%%%%%%%%%%%%%%%%%%%%%%%%%%%%%%%%%%%%%%%%%%
{\it 2).  Confronting the solar and reactor/accelerator results.} 
%%%%%%%%%%%%%%%%%%%%%%%%%%%%%%%%%%%%%%%%%%%%%%%%%%%%%%%%%%%%%%%%%
If non-zero $\sin^2\theta_{13}$ is measured,   
our analysis will allow to understand relevance of 
the 1-3 mixing to the precision solar neutrino studies. 
For fixed $\Delta m^2_{21}$, the 1-3 mixing will change determination of 
$\theta_{12}$. 

Notice that  dependence of sensitivity  of the solar neutrino studies 
to $\theta_{13}$ 
on the mass $\Delta m^2_{31}$ is very weak (via matter corrections to 
1-3 mixing) in contrast to 
laboratory measurements. So,  for small $\Delta m^2_{13}$ the former 
will have some advantage.\\

 If $\theta_{13}$ effect is not found in future laboratory experiments 
and the upper bound at the level $\sin^2\theta_{13} < 0.01$ is 
established,  the effect of 1-3 mixing will be irrelevant  even for 
measurements with the 1\% accuracy. The uncertainty of future 
measurements of $\sin^2\theta_{12}$ due to unknown  
$\theta_{13}$ will be reduced down to 0.01. Furthermore,  the 
precise measurements of $\sin^2\theta_{12}$ by CC/NC will have 
the smallest uncertainty due to larger slope of the iso-lines.\\

The analysis performed in this paper 
will allow us to understand new physics if inconsistencies in results 
of measurements will  show up. 
Implications will depend on character of results, and in the positive 
case, on particular value of $\sin^2\theta_{13}$. 
Comparing the laboratory bound with the results of  solar 
neutrino studies one can realize  new physics  effects  which 
may show up in the solar neutrinos only and  can not be 
found otherwise. Let us consider some examples. 

Suppose that in the standard $3\nu$ analysis 
precision solar neutrino measurements will 
give  a nonzero value of $\sin^2\theta_{13}$ which is 
larger than the upper bound from the reactor/accelerator experiments. 
Such a situation may indicate some additional suppression of the 
solar neutrino flux at low energies. Indeed, this would shift the 
low iso-contours to the left - down.  It would be equivalent of taking the 
low   iso-contours in our plots ({\it e.g.},  in fig.~\ref{ccbyncga}) 
with larger survival probability. 
The intersection with high energy iso-contours will shift down 
and therefore the inferred value of $\sin^2\theta_{13}$ will be reduced. 
This situation can be reproduced in the presence of  sterile neutrino 
with  $\Delta m^2 < 10^{-5}$ eV$^{2}$ \cite{HSs}. In this case 
the extracted value of $\sin^2\theta_{12}$ reduces too.  

Agreement between the solar and laboratory measurements can be also  
achieved by an additional suppression of the high energy (boron) flux
due to flavor conversion, so that CC/NC lines move to the right -  down. 
In this case  also the intersection of the high and low 
iso-contours moves down  and  the inferred value of $\sin^2\theta_{12}$ 
increases. Such an effect can be produced by certain non-standard neutrino 
interactions. 

The  value of $\sin^2\theta_{13}$ extracted from the solar neutrino 
studies can be diminished if both the low and high energy fluxes are 
reduced 
by some additional mechanism. In fact, this imitates the effect of the 
1 - 3 mixing and can be achieved by additional non-standard interactions. 
\\

Let us consider the opposite situation: 
$\sin^2\theta_{13}$ is  discovered  in the reactor/accelerator 
experiments, but the solar neutrino data prefer zero or smaller 
value of $\sin^2\theta_{13}$. 
This can be explained if  the survival probability is larger 
than the one predicted  for given values of 
$\theta_{12}$ and $\theta_{13}$. Again some scenarios 
with modified dynamics of conversion due to new interactions of 
neutrinos can reproduce such a situation.\\

Similarly discrepancy can be realized comparing values of the 1-2 mixing
extracted from the solar neutrino data and reactor/accelerator
experiments.
Indeed, the $\sin^2 \theta_{12}$ can be measured 
with 3\% accuracy at $1\sigma$ (1 d.o.f.) 
in the reactor experiments with baseline  50 - 70 km 
\cite{th12,min,th12new} provided that the 1-3 mixing is determined 
(restricted) accurately enough from independent measurements.    
Since a non-zero $\theta_{13}$ drives the 1-2 mixing to smaller 
values in such an experiment, combination with high energy solar 
neutrino experiments for which non-zero $\theta_{13}$ drives 
$\sin^2\theta_{12}$ to higher values can resolve the 
$\theta_{13}-\theta_{12}$ degeneracy. This is similar to combination 
of low energy and high energy solar neutrino experiments as discussed in the 
present article.  
%If the latter is not done the degeneracy of 1-2 and 1-3 mixings 
%is present  in these reactor experiments which is similar to 
%the  degeneracy in studies of the low energy part of 
%the solar neutrino spectrum (averaged vacuum 
%oscillations driven by 1-3 mixing 
%and vacuum oscillations due to 1 - 2 mixing). 
In this case one can compare the 
($\sin^2 \theta_{12} - \sin^2 \theta_{13}$) allowed regions from 
the solar neutrino studies and from reactor experiments.

%%%%%%%%%%%%%%%%%%%%%%%%%%%%%%%%%%%%%%%%%%%%%%%%%%%%%%%%%
\section{Conclusion}
%%%%%%%%%%%%%%%%%%%%%%%%%%%%%%%%%%%%%%%%%%%%%%%%%%%%%%%

\noindent
1).  We have derived formula for the survival probability in 
the three neutrino context (\ref{pro2}) which includes all  
corrections relevant for the future precision of measurements. 
We estimated the accuracy of approximation made. \\ 

\noindent
2). In the three generation context in the first approximation 
(neglecting the matter effect on $\theta_{13}$)  
the solar neutrino probabilities are functions of the mass split
$\Delta m^2_{21}$  and the mixing  angles:  $\theta_{12}$ 
and $\theta_{13}$. 
The split $\Delta m^2_{21}$ is already determined with a high precision 
by current data. Furthermore,  the survival probability has a weak 
dependence on $\Delta m^2_{21}$ in the allowed region.  
In view of this we have constructed various plots in $\sin^2\theta_{12} -
\sin^2\theta_{13}$ plane which clearly demonstrate the 
degeneracy of these mixing angles and also show how the uncertainty 
in one is going to affect the precision determination of the other. 
Synergy between the high and low energy solar neutrino 
flux measurement provides  a better constraint on $\theta_{13}$ and allows us 
to resolve the degeneracy of these mixing angles.\\  

\noindent
3).  Among the existing  solar neutrino experiments  only the 
Ga-experiment is sensitive  to the low energy part of the spectrum 
and in particular to the  
pp-neutrinos. Therefore we 
construct  the iso-rates for Ga experiments in the $\sin^2\theta_{12} - 
\sin^2\theta_{13}$ plane  and investigate how the combinations of these 
with the iso-rates of  Cl, SK and CC/NC ratio in SNO constrain  
$\theta_{13}$. 

At present the sensitivity of the solar neutrino 
experiments to $\theta_{13}$ is rather weak   
and no indications of non-zero value has been obtained.  
Our analysis of all  solar neutrino data 
leads to statistically insignificant non-zero best fit value: 
$\sin^2 \theta_{13} = 0.01$.  
Similar value, $\sin^2 \theta_{13} \sim  0.007$, follows from  
the fit of the solar neutrino data for 
$\Delta m^2 = 8 \cdot 10^{-5}$ eV$^2$ selected by the KamLAND experiment. 
The combination of the CC/NC ratio at SNO  and  $Q_{Ge}$ rate, 
which are essentially  independent of the solar neutrino fluxes,  gives 
$\sin^2  \theta_{13} = 0.017 \pm 0.026$ ($1\sigma$).

The global fit of all existing  data (including 
reactor, accelerator  and atmospheric neutrino results) leads to  $\sin^2 
\theta_{13} = 0.00$ as 
the best fit point and to the upper bounds $\sin^2 \theta_{13} = 
0.041~(0.061)$
at $90\%~ (3\sigma)$ from the two parameter plots in 
$\sin^2\theta_{12} - \sin^2\theta_{13}$ plane. \\

\noindent
4). The precision of   $\theta_{13}$ determination  will  improve with 
future accurate  measurements  of the pp-neutrino flux,  the CC/NC ratio 
and the  energy spectrum of events  at 
$E > 4$ MeV. We find that measurements with 5\% error in  the 
CC/NC ratio and 2\% error in the pp-neutrino flux in  
$\nu e$ scattering will have a sensitivity ($1\sigma$ error)  $\sin^2 
\theta_{13} \sim 0.015$ 
($1\sigma$). The value $\sin^2 \theta_{13} \sim 0.01$ ($1\sigma$) looks like 
the ultimate sensitivity which could be reached in the solar 
neutrino experiments of the next generation. 

Since the probability involved is $P_{ee}$ the problem of
degeneracy with CP phase $\delta$ does not appear and solar 
neutrinos can provide a clean channel for  $\theta_{13}$ measurements 
like the reactor neutrinos.\\

\noindent
5). It is likely that new  accelerator and reactor results will be obtained 
first. Comparison of the results from solar neutrinos and 
accelerator/reactor experiments may  confirm each other 
and further improve determination of $\sin^2 \theta_{13}$ or reveal 
some  discrepancy which will indicate a new physics beyond 
usual interpretation of the solar neutrino anomaly. 

Determination of $\sin^2 \theta_{13}$ will eliminate or reduce  the 
$\theta_{12} - \theta_{13}$ degeneracy  
thus improving measurements of 
1-2 mixing. Inversely, the better determination of 
$\sin^2\theta_{12}$  can help in improving the 
precision of the $\sin^2\theta_{13}$ measurements in the   solar neutrino 
studies.

Comparison of the results of  $\sin^2\theta_{13}$ 
(as well as $\sin^2\theta_{12}$) measurements  in  solar neutrinos  
and in reactor and accelerator experiments   
may uncover new interesting physics beyond LMA MSW picture which can 
not be found otherwise.

\vskip 8pt
S.G. would like to thank The Abdus Salam International Centre for
Theoretical Physics for hospitality and A. Bandyopadhyay and S. Choubey 
for using some of the codes developed with them. The authors are grateful 
to E. Lisi for valuable comments. 

\vspace{-0.3cm}

%%%%%%%%%%%%%%%%%%%%%%%%%%%%

%%%%%%%%%%%%%%%%%%%%%%%%%%%%


\begin{thebibliography}{}
%%%%%%%%%%%%%%%%%%%%%%%%%%

\bibitem{chooz}
 M.\ Apollonio \textit{et al.},
                 {\em Phys. Lett. }{\bf B466} (1999) 415; 
%``Search for neutrino oscillations on a long base-line at the CHOOZ  
% nuclear power station,''
Eur.\ Phys.\ J.\ C {\bf 27} (2003) 331. 
%%[arXiv:hep-ex/0301017].

\bibitem{palo}
F. Boehm \textit{et al.},
{\em Phys. Rev.} {\bf D62} (2000) 072002.

%\cite{Fogli:2005cq}
%\bibitem{Fogli:2005cq}
\bibitem{lisi05}
  G.~L.~Fogli, E.~Lisi, A.~Marrone and A.~Palazzo,
  %``Global analysis of three-flavor neutrino masses and mixings,''
  arXiv:hep-ph/0506083.
  %%CITATION = HEP-PH 0506083;%%

\bibitem{strumia}
A. Strumia and F. Vissani, arXiv:hep-ph/0503246.  

\bibitem{john}J. N. Bahcall, M.C. Gonzalez-Garcia 
C. Pena-Garay, JHEP {\bf 0408}, 016 (2004);
[hep-ph/0406294]. 
%SOLAR NEUTRINOS BEFORE AND AFTER NEUTRINO 2004.


\bibitem{kl766us}
A.~Bandyopadhyay {\it et al.},
%, S.~Choubey, S.~Goswami, S.~T.~Petcov and D.~P.~Roy,
%``Update of the solar neutrino oscillation analysis with the 766-Ty KamLAND
%spectrum,''
hep-ph/0406328.


\bibitem{valen04}
M. Maltoni, T. Schwetz, M.A. Tortola, J.W.F. Valle, 
New J.
Phys. {\bf 6} 122 (2004),
[hep-ph/0405172] v.3. 
%STATUS OF GLOBAL FITS TO NEUTRINO OSCILLATIONS.

\bibitem{concha04} M. C. Gonzalez-Garcia, [hep-ph/0410030]. 


\bibitem{sgnu04}S. Goswami, talk at Neutrino 2004, Paris, 
http://neutrino2004.in2p3.fr, [hep-ph/0409224]. 

\bibitem{minos}
R.~Saakian,
%``Status of the MINOS experiment,''
Nucl.\ Phys.\ Proc.\ Suppl.\  {\bf 111}, 169 (2002).
%%CITATION = NUPHZ,111,169;%%

\bibitem{cngs}
http://proj-cngs.web.cern.ch/proj-cngs/. 

\bibitem{jparc}
http://jkj.tokai.jaeri.go.jp/.

\bibitem{Huber}
P.~Huber, M.~Lindner, T.~Schwetz and W.~Winter,
%``Reactor neutrino experiments compared to superbeams,''
Nucl.\ Phys.\ B {\bf 665} (2003) 487
[hep-ph/0303232].

\bibitem{2chooz}{\it Letter of Intent for Double-CHOOZ} 
F. Ardellier et. al., hep-ex/0405032. 

\bibitem{yasuda}
%\cite{Minakata:2002jv}
%\bibitem{Minakata:2002jv}
H.~Minakata, H.~Sugiyama, O.~Yasuda, K.~Inoue and F.~Suekane,
%``Reactor measurement of Theta(13) and its complementarity to long-baseline
%experiments,''
Phys.\ Rev.\ D {\bf 68}, 033017 (2003)
[arXiv:hep-ph/0211111];
%%CITATION = HEP-PH 0211111;%%

%\cite{Lunardini:2003eh}
\bibitem{Lunardini:2003eh}
C.~Lunardini and A.~Y.~Smirnov,
%``Probing the neutrino mass hierarchy and the 13-mixing with supernovae,''
JCAP {\bf 0306}, 009 (2003)
[hep-ph/0302033].


%\cite{Fogli:2005ad}
\bibitem{Fogli:2005ad}
  G.~L.~Fogli, E.~Lisi, A.~Mirizzi and D.~Montanino,
  %``Supernova neutrino physics with next-generation water-Cherenkov
  %detectors,''
  Nucl.\ Phys.\ Proc.\ Suppl.\  {\bf 145}, 343 (2005).
  %%CITATION = NUPHZ,145,343;%%



\bibitem{nufac}
 M. Lindner, Int. J. Mod. Phys. A{\bf 18}, 3921, (2003) and references 
therein.   

%%%%%%%%%%solar analysis in 3nu%%%%%%%%%%%%%%%%%%%%%%%%%%%%%%%%%

\bibitem{bari1}G. L. Fogli, E. Lisi, A. Marrone,  D. 
Montanino, A. Palazzo, [hep-ph/0104221]. 
%ATMOSPHERIC, SOLAR, AND CHOOZ NEUTRINOS: A GLOBAL THREE GENERATION 
%ANALYSIS.


\bibitem{bari}G.L. Fogli, G. Lettera, E. Lisi, A. Marrone, A. Palazzo, 
A. Rotunno,  Phys. Rev. {\bf D66}, 093008, (2002), 
G.L. Fogli, E. Lisi, A. Marrone, D. Montanino, 
A. Palazzo, A.M. Rotunno, Phys. Rev. D{\bf 69},  017301, (2004)
[hep-ph/0308055]. 

\bibitem{roadmap} 
J.~N.~Bahcall and C.~Pena-Garay, JHEP {\bf 0311}, 004 (2003)
[hep-ph/0305159].

\bibitem{concha}M. C. Gonzalez-Garcia, C. Pena-Garay,  
Phys. Lett. {\bf B527} 199, (2002). 
%ON THE EFFECT OF THETA (13) ON THE DETERMINATION OF SOLAR OSCILLATION 
%PARAMETERS AT KAMLAND.

\bibitem{carlos}M.C. Gonzalez-Garcia, C. Pena-Garay,
Phys. Rev. D{\bf 68} 093003 (2003). 
%e-Print Archive: hep-ph/0306001
%THREE NEUTRINO MIXING AFTER THE FIRST RESULTS FROM K2K AND KAMLAND.

\bibitem{balan}A. B. Balantekin, H. Yuksel,  
J. Phys. G{\bf 29} 665 (2003). 

\bibitem{valen03}M. Maltoni, T. Schwetz, M.A. Tortola, J.W.F. Valle, Phys. 
Rev. D {\bf 68}, 113010 (2003),  
M.C. Gonzalez-Garcia, M. Maltoni, C. Pena-Garay, J.W.F. Valle,  
Phys. Rev. D {\bf 63}  033005 (2001). 

\bibitem{HS03} P. C. de Holanda, A.Yu. Smirnov, 
Astropart. Phys. {\bf21}, 287, (2004) [hep-ph/0309299]. 

\bibitem{HS02}
P. C. de Holanda, A.Yu. Smirnov,   
JCAP {\bf 0302}, 001, (2003);  
P.C. de Holanda, A.Yu. Smirnov, Phys. Rev. D{\bf66}, 113005, (2002). 

\bibitem{kkthree}
A.~Bandyopadhyay, S.~Choubey, S.~Goswami and K.~Kar,
%``Three generation neutrino oscillation parameters after SNO,''
Phys.\ Rev.\ D {\bf 65} (2002) 073031
[hep-ph/0110307]. 
%%CITATION = HEP-PH 0110307;%%

\bibitem{kl162us}
%\cite{Bandyopadhyay:2002mc}
%\bibitem{Bandyopadhyay:2002mc}
A.~Bandyopadhyay, S.~Choubey, R.~Gandhi, S.~Goswami and D.~P.~Roy,
%``Testing the solar LMA region with KamLAND data,''
J.\ Phys.\ G {\bf 29}, 2465 (2003)
[hep-ph/0211266].
%%CITATION = HEP-PH 0211266;%%
%
%\bibitem{solfit1}
A.~Bandyopadhyay, S.~Choubey, R.~Gandhi, S.~Goswami and D.~P.~Roy,
Phys.\ Lett.\ B {\bf 559}, 121 (2003) [arXiv:hep-ph/0212146].


\bibitem{ind2}
A. Bandyopadhyay, S. Choubey, S. Goswami, S.T. Petcov, D.P. Roy,  
Phys. Lett. B{\bf 583}, 134 (2004). 


\bibitem{3nu}S. C. Lim, In Proc. of BNL Neutrino Workshop, Upton, NY, 
USA 1987, edited by M. Murtagh, BNL-52079, C87/02/05.  

\bibitem{3nuS}
X. Shi, D. N. Schramm, Phys. Lett. B {\bf 283}, 305 (1992), 
X. Shi, D.N. Schramm, J. N. Bahcall,  
Phys. Rev. Lett. {\bf 69}, 717 (1992). 

\bibitem{lisi}G. L. Fogli, E. Lisi, A. Palazzo,  
Phys. Rev. D {\bf 65}, 073019 (2002),  
[hep-ph/0105080].  

\bibitem{lim03} C.S. Lim, K. Ogure, H. Tsujimoto, 
Phys. Rev. D{\bf 67}, 033007 (2003), [hep-ph/0210066].

%\bibitem{ours} A. Yu. Smirnov, in Proc. of the Int. Symposium on Neutrino 
%Astrophysics, Frontiers of  Neutrino Astrophysics,   
%October 19 - 22,  Takayama, Japan 1992, Edited by Y. Suzuki and K. 
%Nakamura, p. 105.    

\bibitem{BOS}
M.~Blennow, T.~Ohlsson and H.~Snellman,
%``Day-night effect in solar neutrino oscillations with three flavors,''
Phys.\ Rev.\ D{\bf 69}, 073006 (2004),  
[hep-ph/0311098].
%%CITATION = HEP-PH 0311098;%%

\bibitem{akh}E. Kh. Akhmedov, M. A. Tortola, J.W.F. Valle,  
JHEP {\bf 0405},  057 (2004), [hep-ph/0404083].

\bibitem{ms}
S. P. Mikheyev and A. Yu. Smirnov, Yad. Fiz. {\bf 42}, 1441 (1985) [
Sov. J. Nucl. Phys. {\bf 42}, 913 (1985)]; Nuovo Cim. {\bf C9}, 17
(1986); S. P. Mikheyev and A. Yu. Smirnov, ZHETF, {\bf 91}, (1986),
[Sov. Phys. JETP, {\bf 64}, 4 (1986)] (reprinted in "Solar neutrinos:
the first thirty years", Eds. J.N.Bahcall {\it et. al.}).

\bibitem{messiah}
 A. Messiah, in Proceedings of the $6$th Moriond
Workshop On Massive Neutrino in Particle Physics and Astrophysics,
ed. O. Fackler and J. Tran Thanh Van, (1986) p.373. 

\bibitem{parke}
W. Haxton, Phys. Rev. Lett. {\bf 57} (1986) 1271;  
S. J. Parke, Phys. Rev. Lett. {\bf 57}(1986) 1275. 

\bibitem{smir04} 
P.~C.~de Holanda, W.~Liao and A.~Y.~Smirnov,
Nucl. Phys. B{\bf 702},   307, (2004), [hep-ph/0404042].

\bibitem{araic}
  A.~N.~Ioannisian and A.~Y.~Smirnov,
  %``Neutrino oscillations in low density medium,''
  Phys.\ Rev.\ Lett.\  {\bf 93} (2004) 241801. 


\bibitem{GPS}
  M.~C.~Gonzalez-Garcia, C.~Pena-Garay and A.~Y.~Smirnov,
%``Zenith angle distributions at Super-Kamiokande and SNO and the 
%solution  of
%the solar neutrino problem,''
  Phys.\ Rev.\ D {\bf 63} (2001) 113004.
 


%\cite{Ioannisian:2004vv}
%\bibitem{ioa}
%A.~N.~Ioannisian, N.~A.~Kazarian, A.~Y.~Smirnov and D.~Wyler,
%arXiv:hep-ph/0407138.

\bibitem{th12new}
%\cite{Bandyopadhyay:2004cp}
%\bibitem{Bandyopadhyay:2004cp}
A.~Bandyopadhyay, S.~Choubey, S.~Goswami and S.~T.~Petcov,
%``High precision measurements of Theta(solar) in solar and reactor neutrino
%experiments,''
[hep-ph/0410283].
%%CITATION = HEP-PH 0410283;%%

\bibitem{kearns}E. Kearns, talk at Neutrino 2004, Paris,
http://neutrino2004.in2p3.fr.


%\bibitem{kl766}
%\cite{:2004mb}
%\bibitem{:2004mb}
%G. Gratta, talk at Neutrino 2004;
%T. Araki {\it et al.},
%  [KamLAND Collaboration],
%``Measurement of neutrino oscillation with KamLAND: Evidence of spectral
%distortion,''
%hep-ex/0406035.
%%CITATION = HEP-EX 0406035;%%


\bibitem{v3}
T. Araki {\it et al.}, [KamLAND Collaboration],
hep-ex/0406035 (v3).
%%CITATION = HEP-EX 0406035;%%






%\cite{Cleveland:nv}
%\bibitem{Cleveland:nv}
\bibitem{cl}
B.~T.~Cleveland {\it et al.},
%``Measurement Of The Solar Electron Neutrino Flux With The Homestake  Chlorine
Detector,''
Astrophys.\ J.\  {\bf 496}, 505 (1998).
%%CITATION = ASJOA,496,505;%%
\bibitem{ga}
%\bibitem{Abdurashitov:2002nt}
J.~N.~Abdurashitov {\it et al.}  [SAGE Collaboration],
% ``Measurement of the solar neutrino capture rate by the Russian-American
%gallium solar neutrino experiment during one half of the 22-year cycle  of
%solar activity,''
J.\ Exp.\ Theor.\ Phys.\  {\bf 95}, 181 (2002)
[Zh.\ Eksp.\ Teor.\ Fiz.\  {\bf 122}, 211 (2002)]
[arXiv:astro-ph/0204245].
%%CITATION = ASTRO-PH 0204245;%%%\cite{Hampel:1998xg}
%\bibitem{Hampel:1998xg}
W.~Hampel {\it et al.}  [GALLEX Collaboration],
%``GALLEX solar neutrino observations: Results for GALLEX IV,''
Phys.\ Lett.\ B {\bf 447}, 127 (1999)
%%CITATION = PHLTA,B447,127;%%
;
%\cite{Altmann:2005ix}
%\bibitem{Altmann:2005ix}
  M.~Altmann {\it et al.}  [GNO Collaboration],
  %``Complete results for five years of GNO solar neutrino observations,''
  Phys.\ Lett.\ B {\bf 616} (2005) 174.
%  [arXiv:hep-ex/0504037].
%%CITATION = HEP-EX 0504037;%%



%\bibitem{Fukuda:2002pe}
\bibitem{sksolar}
S.~Fukuda {\it et al.}  [Super-Kamiokande Collaboration],
%``Determination of solar neutrino oscillation parameters using 1496 days  of Su per-Kamiokande-I data,''
Phys.\ Lett.\ B {\bf 539}, 179 (2002), 
[hep-ex/0205075].
%%CITATION = HEP-EX 0205075;%%

\bibitem{sno}%\cite{Ahmad:2002ka}
%\bibitem{Ahmad:2002ka}
Q.~R.~Ahmad {\it et al.}  [SNO Collaboration],
Phys.\ Rev.\ Lett.\  {\bf 89}, 011302 (2002)
[nucl-ex/0204009].
%%CITATION = NUCL-EX 0204009;%%\bibitem{Ahmed:2003kj}
S.~N.~Ahmed {\it et al.}  [SNO Collaboration],
%``Measurement of the total active B-8 solar
% neutrino flux at the Sudbury Neutrino
% Observatory with enhanced neutral current sensitivity,''
Phys.\ Rev.\ Lett.\  {\bf 92}, 181301 (2004)
[arXiv:nucl-ex/0309004].
%%CITATION = NUCL-EX 0309004;%%

%\cite{Aharmim:2005gt}
%\bibitem{Aharmim:2005gt}
\bibitem{salt2}
  B.~Aharmim {\it et al.}  [SNO Collaboration],
  %``Electron energy spectra, fluxes, and day-night asymmetries of B-8 solar
  %neutrinos from the 391-day salt phase SNO data set,''
  arXiv:nucl-ex/0502021.
  %%CITATION = NUCL-EX 0502021;%%


\bibitem{bp04}
%\cite{Bahcall:2004fg}
%\bibitem{Bahcall:2004fg}
J.~N.~Bahcall and M.~H.~Pinsonneault,
%``What do we (not) know theoretically about solar neutrino fluxes?,''
Phys.\ Rev.\ Lett.\  {\bf 92}, 121301 (2004)
[astro-ph/0402114].
%%CITATION = ASTRO-PH 0402114;%%

%\cite{Bahcall:2005dd}
\bibitem{bahcall05}
  J.~N.~Bahcall, S.~Basu and A.~M.~Serenelli,
%``What Is The Neon Abundance Of The Sun?,''
arXiv:astro-ph/0502563.

\bibitem{sno3}
K.  Graham, talk at NOON 2004, February 11-15, 2004,
Tokyo, Japan,
{\it http://www-sk.icrr.u-tokyo.ac.jp/noon2004/};
H. Robertson for the SNO Collaboration, Talk given at TAUP 2003,
Univ. of Washington, Seattle, Washington, September 5 - 9, 2003,
%transparencies at:
{\it http://mocha.phys.washington.edu/taup2003}

\bibitem{borexnu2004}C. Galbiati, talk at Neutrino 2004, Paris, 
http://neutrino2004.in2p3.fr.

\bibitem{ppex}%\bibitem{XMASS}
M. Nakahata,
                 Talk given at the Int. Workshop on Neutrino
                 Oscillations and their Origin
                 (NOON2004), February 11 - 15, 2004, Tokyo, Japan;
S. Sch\"{o}nert, talk at Neutrino 2002, Munich, Germany,

\bibitem{lownu}see {\it e.g.},  Y. Suzuki, 
Nucl. Phys. B (Proc. Suppl.) {\bf 143}, 27 (2005)
``Neutrino 2004'' 

\bibitem{hyperk}T. Nakaya, 
Nucl. Phys. B (Proc. Suppl.) {\bf 118}, 210 (2003)
``Neutrino-2002'' 
%Next-Generation Cherenkov Detector Hyper-Kamiokande,
K. Nakamura, talk at the workshop ``Neutrinos and Implications for Physics 
Beyond the Standard Model'' 2002,
http://insti.physics.sunysb.edu/itp/conf/neutrino/talks/nakamura.pdf. 

\bibitem{UNO}UNO Proto-collaboration, UNO Whitepaper:
Physics Potential and Feasibility of UNO, SBHEP-01-03(2000),
http://nngroup.physics.sunysb.edu/uno/;
see also C. K. Jung 2002, hep-ex/0005046. 

\bibitem{barger} 
%\cite{Barger:2001yr}
%\bibitem{Barger:2001yr}
V.~Barger, D.~Marfatia and K.~Whisnant,
%``Breaking eight-fold degeneracies in neutrino CP violation, mixing, and  mass
%hierarchy,''
Phys.\ Rev.\ D {\bf 65}, 073023 (2002)
[hep-ph/0112119].
%%CITATION = HEP-PH 0112119;%%

\bibitem{HSs}
%\cite{deHolanda:2003tx}
%\bibitem{deHolanda:2003tx}
P.~C.~de Holanda and A.~Y.~Smirnov,
%``Homestake result, sterile neutrinos and low energy solar neutrino
%experiments,''
Phys.\ Rev.\ D {\bf 69}, 113002 (2004)
[hep-ph/0307266].
%%CITATION = HEP-PH 0307266;%%


\bibitem{th12} A.~Bandyopadhyay, S.~Choubey and S.~Goswami,
Phys.\ Rev.\ D {\bf 67}, 113011 (2003). 
 

\bibitem{min} 
H.~Minakata, H.~Nunokawa, W.~J.~C.~Teves and R.~Zukanovich Funchal, 
[hep-ph/0407326]. 



%%%%%%%%%%%%%%%%%%%%%%%%%%%%
\end{thebibliography}
\end{document}